%% file: main.tex
\documentclass[letterpaper,twocolumn,10pt]{article}

\usepackage{usenix-2020-09}
\usepackage[utf8]{inputenc}
\usepackage[T1]{fontenc}
\usepackage{graphicx}
\usepackage{booktabs}
\usepackage{amsmath}
\usepackage{amsfonts}
\usepackage{microtype}
\usepackage{caption}
\usepackage{placeins}
\usepackage{algorithm}
\usepackage{algorithmic}
\usepackage{etoolbox}

\newcommand{\systemname}{Talaria}

\makeatletter
\patchcmd{\@maketitle}{\vbox to 2.5in}{\vbox to 2.2in}{}{}
\makeatother

\title{\Large \bf Talaria: Session-Aware Serverless Serving of Hundred-Billion-Parameter LLMs}

\author{
  \begin{tabular*}{\dimexpr\textwidth-2\tabcolsep\relax}{@{\extracolsep{\fill}}ccccc@{}}
    {\normalfont Utopia Meng} & {\normalfont Unicornt Zhao} &
    {\normalfont Derek Li} & {\normalfont Goalen Gao} &
    {\normalfont Frank Du} \\
    {\normalfont\footnotesize 2197096044@qq.com} &
    {\normalfont\footnotesize unicornt@unicornt.xyz} &
    {\normalfont\footnotesize 940619089@qq.com} &
    {\normalfont\footnotesize 94205123@qq.com} &
    {\normalfont\footnotesize 1093782566@qq.com}
  \end{tabular*}
}

\begin{document}

\maketitle

\input{sections/abstract}
\input{sections/01-introduction}
\input{sections/02-background}
\input{sections/03-method}
\input{sections/04-implementation}
\input{sections/05-evaluation}
\FloatBarrier
\input{sections/06-related-work}
\input{sections/07-conclusion}

\bibliographystyle{unsrt}
\bibliography{ref}

\appendix
\input{sections/appendix}

\end{document}

%% file: sections/abstract.tex
\begin{abstract}
Serverless multi-model LLM systems multiplex popularity-skewed model catalogs
over shared GPU pools, yet typically schedule each request independently.
Tool-using agents break this abstraction: a session repeatedly calls an LLM
across short tool gaps, carries a long reusable KV prefix, and is judged by
session completion time (SCT). Load-only routing can separate a continuation
from both its model and KV state, while round-based model multiplexing can delay
even a correctly placed continuation until the target model's next slot. Both
failures are especially costly for hundred-billion-parameter models: their
weights constrain residency, while long-context KV is expensive to reconstruct
or move.

We present \systemname{}, a session-aware serverless multi-model serving system
that makes session continuity a joint placement-and-admission decision. Its
router ranks placements by model residency, KV locality, and instance pressure,
while soft reservations account for likely returns in the last serving
instance's admission budget. Session-prefill (SP) admits budget-eligible
continuations before the active model slot closes. An instance-local substrate
keeps HBM addresses stable, preserves host-restorable KV, and stages weights
across model switches.

On a single TP=8 server, we replay 30 SWE-Bench model-sessions (960 calls) over
three models, each with more than 100B total parameters. Against an otherwise
identical round scheduler with SP, host-KV restoration, and D2D staging
disabled, \systemname{} cuts p50 SCT from 1000\,s to 189\,s and p95 from
2296\,s to 867\,s, speedups of 5.3$\times$ and 2.6$\times$.
\end{abstract}

%% file: sections/01-introduction.tex
\section{Introduction}
\label{sec:intro}

Modern LLM serving platforms host popularity-skewed model catalogs on shared
GPU clusters: a few hot models carry stable traffic, while a long tail is
invoked only intermittently~\cite{stanford2025aiindex,ghosh2026hfopensource,
sheng2024slora,wu2024dlora,xiang2025aegaeon}. Statically reserving GPUs for
every model wastes accelerator capacity on the tail, motivating serverless
multi-model serving: demand-driven opening, closing, and time-sharing of models
over pooled GPUs. ServerlessLLM optimizes cluster placement and cold-start
loading~\cite{fu2024serverlessllm}; AlpaServe, MuxServe, and Aegaeon multiplex
or fast-switch models on pooled resources~\cite{li2023alpaserve,duan2024muxserve,
xiang2025aegaeon}. These techniques make model residency a first-class concern,
but their placement and scheduling abstractions do not explicitly preserve
continuity across calls in an agent session. Large models intensify this gap:
when the aggregate working set exceeds HBM capacity, a placement miss can
trigger weight loading together with KV restoration or prefix recomputation.

Tool-using agents expose why a request alone is an insufficient scheduling unit.
A user task alternates between model calls and tool execution, carrying state
from one call to the next~\cite{yao2023react,schick2023toolformer,
yang2024sweagent,luo2025autellix}.
Across 445 SWE-Bench Verified sessions~\cite{jimenez2024swebench}
(7{,}386 calls), median session depth is 11~model calls (p95: 49).
Prompt-length measurements ($n=5{,}697$) have a median of 25K~tokens (p95:
105K), and 96.8\% of measured continuations ($n=5{,}270$) reuse more than half
of the prior call's prefix. The median tool-use gap is only 0.39\,s---over an
order of magnitude shorter than the 4.3\,s needed to recompute a median-length
prefix on Qwen3-235B/TP=8. These timescales create a continuity window:
continuations return quickly, so preserving their large, mostly reusable
prefixes can avoid seconds of recomputation. Treating each return as a fresh
request repeatedly incurs avoidable state-recovery cost across a session.

\begin{figure*}[!t]
  \centering
  \includegraphics[width=\textwidth]{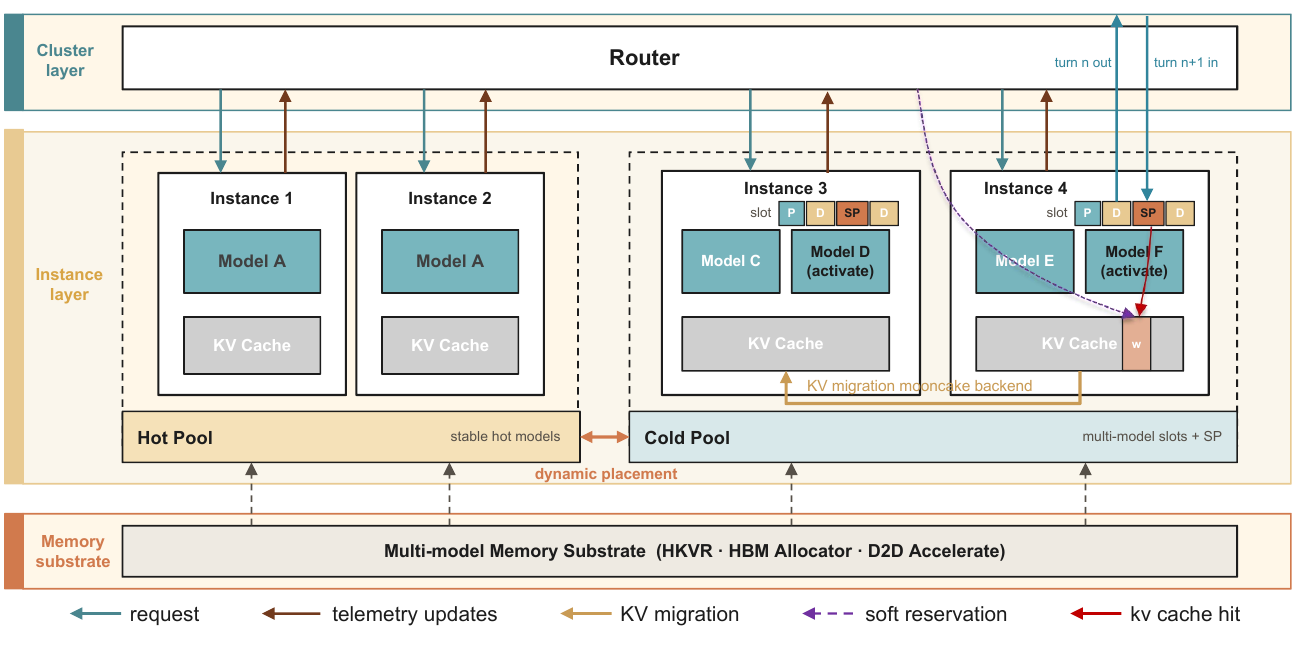}
  \caption{\systemname{} architecture. The router preserves return affinity
  across hot and cold pools; SP admits eligible mid-slot returns, while each
  cold-pool instance preserves restorable KV across model switches.}
  \label{fig:system-overview}
\end{figure*}

Per-call scheduling therefore misaligns with session completion time along two
coupled dimensions.
\textbf{Spatial mismatch:} a router using only load or queue length may send
a continuation to the least-loaded instance, which may hold neither the target
weights nor the session KV; for long-context sessions, KV movement or prefix
recomputation can exceed the queueing delay it avoids.
\textbf{Temporal mismatch:} even when routing is correct, a multi-model
scheduler that rotates through models in fixed-length
rounds~\cite{xiang2025aegaeon} may defer a returning session until the next
slot for that model---KV locality is preserved spatially but squandered on the
time axis.
Under prefill/decode disaggregation, the same locality problem spans tiers:
the returning session enters through a prefill path, while reusable KV is
valuable on the decode side.

These failure modes compound. KV-aware routing can preserve the right state yet
still leave a continuation waiting until the target model's next slot;
mid-slot admission can exploit preserved state only when routing has placed the
continuation on the right instance. The cluster layer must land a return on a
useful instance, and the instance layer must expose a timely execution
opportunity.

We present \systemname{}, a serverless multi-model LLM inference system for
agentic workloads (Figure~\ref{fig:system-overview}). \systemname{} realizes
session continuity through coordinated placement, admission, and state
restoration. At deployment time, it configures a hot pool of pinned
single-model instances and a cold pool whose instances token-level time-share a
set of long-tail models. A session-aware router uses soft reservations to
account for likely returns in the serving instance's next eligible model-slot
budget (\S\ref{sec:router}); calls without a usable reservation are placed by
ranking candidate instances according to model residency, KV residency, and
pressure.
Within each cold-pool slot, co-located prefill/decode execution enables
session-prefill (SP) to admit eligible returns mid-slot, reusing device-resident
KV or restoring host-restorable KV while the model remains active. A
per-instance memory substrate coordinates HBM allocation, host-side KV state,
and weight staging to reduce the switching and restoration costs charged to
each round.

On a single TP=8 server, we replay ten SWE-Bench issues across Qwen3-235B,
GLM5-nvfp4, and Qwen3.5-122B-A10B, yielding 30 model-sessions and 960 calls.
For each model-session, fixed replay holds request bodies, model and session
IDs, per-session call order, completion-token counts, and return gaps constant
across configurations. This prevents agent-path divergence; global interleaving
may still differ because policies change completion times.
Against an otherwise identical round scheduler with SP, host-KV restoration,
and D2D staging disabled, \systemname{}'s instance-local mechanisms cut p50
session completion time from 1000\,s to 189\,s and p95 from 2296\,s to
867\,s---speedups of 5.3$\times$ and 2.6$\times$. On a separate two-worker
testbed, residency-aware routing reduces high-load avoidable model opens from
37 to 1 and TTFT p95 from 8.07\,s to 5.28\,s versus least-pressure routing.

This paper makes three contributions.
\begin{itemize}
  \setlength{\topsep}{2pt}
  \setlength{\itemsep}{2pt}
  \setlength{\parsep}{0pt}
  \item \textbf{Session continuity as a scheduling abstraction.} We identify
  session continuity as the missing unit in serverless multi-model serving for
  agents. From 445 agent sessions, we quantify the depth, return timing, and
  reusable state that make request-level scheduling costly, then isolate
  spatial placement and temporal admission with controlled experiments.
  \item \textbf{Joint session-continuity scheduling.} At the cluster layer,
  soft reservation accounts for likely returns without pinning device KV; at
  the instance layer, SP provides budgeted mid-slot admission. The two
  mechanisms address complementary failure modes along the same return path.
  \item \textbf{Switch-resilient instance substrate.} A stable HBM layout and
  host-restorable KV preserve usable session state across model switches;
  opportunistic D2D staging \mbox{reduces} the switching cost charged to each round.
\end{itemize}

%% file: sections/02-background.tex
\section{Background and Motivation}
\label{sec:background}

We first characterize agent sessions in a real agent trace (\S\ref{sec:bg-session}),
then show how request-level placement loses spatial locality
(\S\ref{sec:bg-placement}) and how round-based multi-model scheduling loses
temporal locality (\S\ref{sec:bg-temporal}). We close with the
memory-management requirements that make session-aware scheduling
implementable (\S\ref{sec:bg-substrate}).

\subsection{Agent sessions are stateful}
\label{sec:bg-session}

\begin{figure}[!t]
  \centering
  \includegraphics[width=\columnwidth]{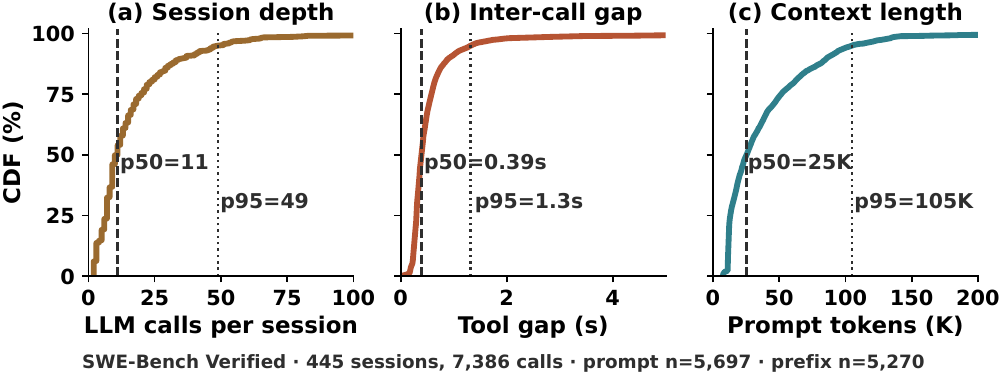}
  \caption{SWE-Bench Verified sessions are deep (median 11 calls),
  fast-returning (median gap 0.39\,s), and long-context (median 25K tokens).
  Axes truncate at 100 calls, 5\,s, and 200K tokens; the gap CDF excludes five
  ${>}30$\,s environment-startup intervals.}
  \label{fig:session-cdf}
\end{figure}

We define an agent \emph{session} as a single user task: the agent issues a
sequence of model calls interleaved with tool invocations until the task
completes. This request pattern follows the agent execution model introduced by
tool-use and software-engineering agents~\cite{yao2023react,
schick2023toolformer,yang2024sweagent}. We characterize 445 completed agent
executions of SWE-Bench Verified tasks~\cite{jimenez2024swebench}, totaling
7{,}386 LLM calls. Prompt length is available for 5{,}697 calls and
reusable-prefix overlap for 5{,}270 continuations. We combine these traces with
measured Qwen3-235B and GLM5-nvfp4 profiles to quantify recovery cost.
Figure~\ref{fig:session-cdf} summarizes session depth, tool gaps, and prompt
lengths.

\textbf{Session depth.}
The median session issues \textbf{11} model calls, with a p95 of 49 and a
maximum of 187. A request-level scheduler therefore sees one user task as many
unrelated arrivals, hiding the prefix reuse and timing dependence between
consecutive calls.

\textbf{Short inter-call gaps.}
The median tool-use gap is only \textbf{0.39\,s}, with a p95 of 1.32\,s.
This is more than ten times shorter than recomputing the median-length prefix,
as measured below. A return therefore often arrives while its prior state
remains worth preserving, whether it is still in HBM or restorable from host
memory.

\textbf{Continuation inputs.}
Prompts are long and highly repetitive: the median call has \textbf{25K}
prompt tokens, the p95 reaches 105K, and 96.8\% of measured continuations reuse
more than half of the previous call's prefix. Recomputing a 25K-token prefix on
Qwen3-235B/TP=8 costs \textbf{4.3\,s} TTFT at p50; at 100K tokens, it reaches
10.9\,s. A single misplaced return can therefore dominate the latency of an
entire session turn.

Taken together, these properties mean that agent serving is not scheduling a
stream of independent requests. It is scheduling a \emph{stateful session}
whose KV state, target model, and return timing persist across calls and must
be tracked explicitly by the serving system.

\subsection{Spatial mismatch: routing to the wrong instance}
\label{sec:bg-placement}

Current serverless routers schedule at request granularity, commonly using
load, queue length, or model-load cost as the routing signal~\cite{
fu2024serverlessllm,xiang2025aegaeon}. This is a poor proxy for a returning
agent call because the fastest instance for the cluster may be the slowest
instance for that session.

Consider two instances in the same cold pool. \textbf{Instance~A} is currently
serving the target model and still holds the session's reusable KV prefix, but
has a short queue. \textbf{Instance~B} is idle, but may hold neither the target
model nor the session KV. A least-load or round-robin router sends the returning
call to~B because its queue is shorter. For this session, B must first make the
target model resident if needed, then reconstruct the prefix. The recomputation
alone costs \textbf{4.3\,s} TTFT on Qwen3-235B/TP=8.

Instance~A may incur queueing or wait until the target model's next slot,
whereas B must additionally activate the model and recover the prefix. Because
recovery cost grows with context length, a shorter queue on B does not imply a
faster return. A load-only router cannot see this asymmetry.

KV-aware routing~\cite{gao2024cachedattention,qin2025mooncake,srivatsa2024preble}
narrows the gap, but it is still incomplete for multi-model serving. Knowing
where the session KV lives is not enough if the router cannot also tell whether
the target model is resident and whether the instance can accept a prefill soon.
Good placement therefore requires joint visibility into model residency, KV
residency, and instance pressure. This visibility must also be session-scoped:
the router must remember which instance just served a session and treat the
next call as a likely continuation, not as an unrelated arrival.

\subsection{Temporal mismatch: admitting at the wrong time}
\label{sec:bg-temporal}

Correct placement is not sufficient if the instance scheduler exposes no time
window in which the returning session can run. A cold-pool instance that
time-shares models in fixed rounds may preserve the session KV in space while
failing to use it in time.

Consider a round-based token-level scheduler such as
Aegaeon~\cite{xiang2025aegaeon}. The decode side commits to one active model
for a slot. After a session finishes a decode step and enters a tool call, the
scheduler may move on to another model.
When the tool returns 0.39\,s later, the session's KV can still be resident in
HBM, but its model window may have closed. The session then waits for that
model's next slot---up to one round---before it can resume.

Systems that disaggregate prefill and decode~\cite{patel2024splitwise,
zhong2024distserve,hu2024tetriinfer} expose an analogous cross-tier mismatch:
the return enters through a prefill path while its reusable state is valuable
on the decode side. \systemname{} focuses on co-located prefill/decode; extending
its admission contract across tiers requires coordinated KV ownership
(\S\ref{sec:conclusion}).

The failure is temporal rather than spatial: the state may be in the right
place, but the scheduler cannot admit the session while it is still useful.
Request-level autoscaling does not solve this problem because it requeues the
returning session as a fresh request, losing the continuity established by the
router. What is needed is mid-slot admission: a returning session should be
able to re-enter the current model window and reuse preserved KV before the
scheduler rotates away.

\subsection{Multi-model memory management}
\label{sec:bg-substrate}

The router and scheduler above assume that model and KV state can be moved,
kept valid, and restored within a bounded time. Existing LLM serving systems
already show that KV layout, paging, and virtual memory are central to serving
performance~\cite{kwon2023vllm,prabhu2024vattention,sheng2023flexgen}. In
large multi-model serving, the requirement is tighter: freeing HBM for one
model can invalidate reusable KV, while slow weight movement consumes the same
round budget that mid-slot admission relies on. Memory management becomes part
of the scheduling substrate.

\textbf{HBM capacity.}
Weights are only part of a model's device footprint. Each model also brings
runtime structures such as CUDA graphs, FlashInfer workspaces, and KV pool
regions. If each model allocates these regions independently, the footprints
accumulate even when request load is low, leaving too little HBM for useful KV
or for staging the next model.

\textbf{Host-side KV consistency.}
Cold-pool switching often requires session KV to leave HBM and later return.
The system must know which model owns each KV block, whether the block is still
valid, and whether the current copy is on device or host. Without this metadata,
the serving stack can restore stale KV and produce incorrect output, or miss a
valid host-side copy and recompute a long prefix unnecessarily.

\textbf{Switch latency.}
Model switching directly consumes the round budget that the scheduler relies
on for TTFT and TPOT. A large-model switch may include weight loading, kernel
reinitialization, and KV-pool reorganization, each of which can be second-scale.
In our fixed-replay measurements, a logical TP=8 H2D-only switch has a 1.50\,s
p50 and a 1.52\,s mean (\S\ref{sec:eval-substrate}). This consumes a substantial
fraction of a 10\,s TTFT budget. If switching
occupies a large fraction of the round, time-sharing stops being a multiplexing
benefit and becomes a throughput tax.

These constraints are coupled. Bounded switching requires spare HBM for staging;
KV restoration requires host-side consistency; and both must complete quickly
enough for mid-slot admission to matter. The memory layer must therefore manage
device memory, host KV state, and weight movement as one substrate rather than
as independent mechanisms.

%% file: sections/03-method.tex
\section{Design}
\label{sec:design}

\subsection{System Overview}

At a high level, \systemname{} separates multi-model agent serving into a
control plane and a data plane.  The control plane---a unified router---decides
where each session call should land.  The data plane---a set of hot and cold
inference instances---executes those calls while maintaining the local
budget and telemetry contracts that the router's decisions assume.

The router creates a soft reservation when a completed call enters tool
execution. It records the serving instance, model, and reusable-prefix handle
and charges that instance's admission budget for a likely return. If the session
returns before the lease expires and the target remains healthy, the router
sends it back to that instance, where SP can admit it while the KV is
device-resident or host-restorable. Requests without a usable reservation take
the normal cost path, which weighs instance pressure, KV recovery, and model
activation. The router never touches token-level execution.

Instances are split into a hot pool and a cold pool.  Hot-pool instances obey a
single invariant: each pins one model and never switches, eliminating switching
tail latency for stable traffic.  Cold-pool instances time-share multiple
long-tail models using round-based scheduling.  Each model slot has the
structure $P \rightarrow D_1 \rightarrow (SP \rightarrow D)^*
\rightarrow D_{\text{fin}}$: queued prefill, initial decode, zero or more
interleaved SP (session-prefill) / decode pairs, and final decode.
Each SP step batch-admits eligible returns that fit the remaining prefill
budget, enabling preserved-prefix reuse without a model switch; excess returns
remain queued for a later opportunity.

Beneath the instance scheduler, a multi-model memory substrate manages the HBM
and host memory of each inference instance, addressing the layout, host-KV, and
switch-latency constraints identified in \S\ref{sec:bg-substrate}. It provides
a stable HBM layout, persists session KV across model switches, and reduces
switch cost through weight staging. The router observes this substrate through
telemetry such as memory pressure and staging slack; the instance scheduler uses
it to restore KV and switch models within the round budget.

One ownership boundary holds across the design: the router reasons about
cluster placement but never executes tokens, while each instance enforces local
execution and memory budgets without reconstructing cluster-wide demand. Pinned
hot instances avoid the switch path; cold instances expose the telemetry and
actuation needed for multiplexing.

\subsection{Configured Hot-Cold Pool Organization}
\label{sec:pooling}

Hot-pool instances pin a single model and never switch; this eliminates switching
tail latency on stable, high-throughput traffic.  Cold-pool instances time-share
long-tail models through round-based scheduling (\S\ref{sec:scheduler}).

Our prototype configures pool membership at deployment time. Within the cold
pool, the router prefers instances that already hold the requested model and
session prefix, subject to admission and HBM constraints. This concentrates
demand on useful state and avoids needless model opens; if the preferred
instance fails admission, cost ranking selects another feasible cold instance
or invokes the configured overload policy. These are per-call placement
decisions and do not reclassify instances.

This configured split separates stable high-demand service from the
multi-model tail while keeping the continuity mechanisms independent of a
particular pool-sizing controller. The telemetry used for placement could also
drive promotion and demotion, but dynamic pool reallocation is outside the
current prototype and evaluation.

\subsection{Session-Aware Residency Routing}
\label{sec:router}

\begin{figure}[t]
  \centering
  \includegraphics[width=\columnwidth]{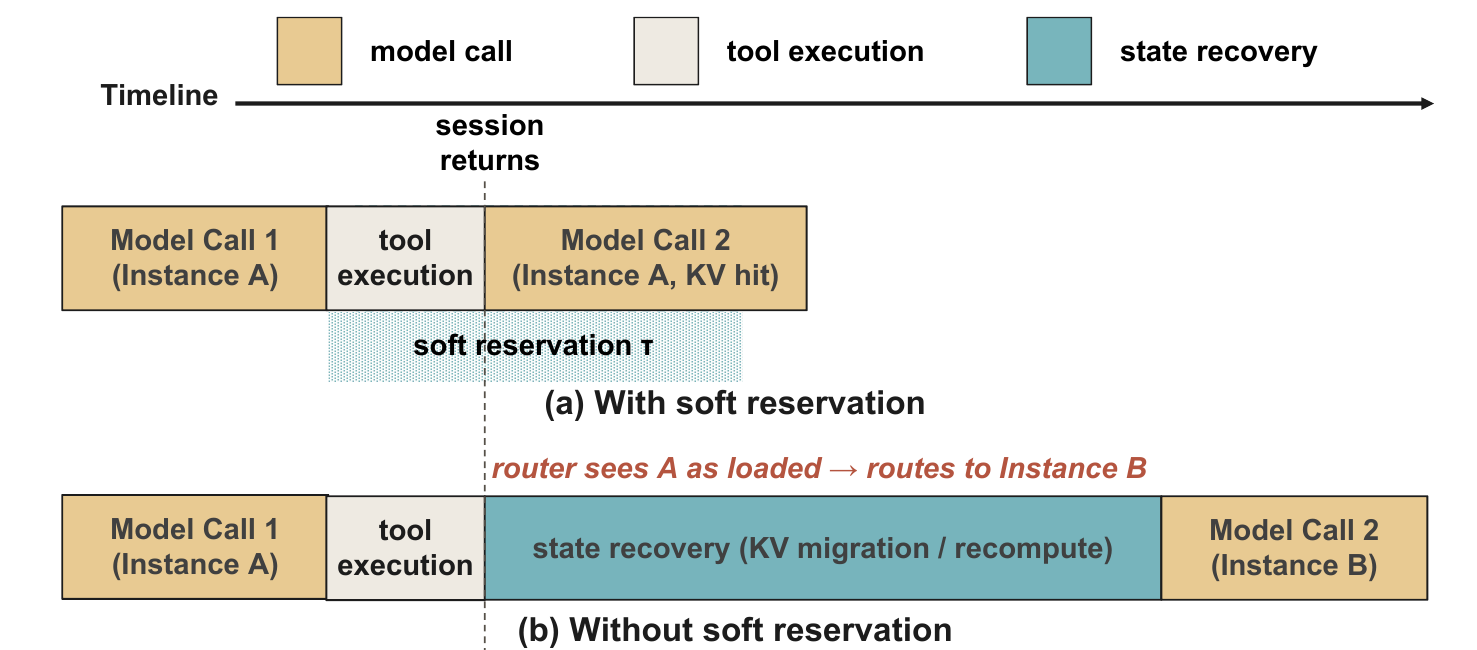}
  \caption{Soft reservation preserves return affinity to an instance with
  device-resident or host-restorable state; it does not pin device KV.}
  \label{fig:soft-reservation}
\end{figure}

The router assigns each session call to an inference instance with two
objectives: preserve KV locality from the previous call, and respect model
and memory capacity constraints.
The two objectives have different time horizons. A session prefix is reusable
when a model call completes, but it may move to host or be evicted while the
session executes a tool. Preserving or restoring that locality is therefore
time-sensitive.
Capacity constraints depend on model residency and aggregate round pressure,
which evolve over scheduling windows rather than a single routing decision.

\paragraph{Soft reservation for returning sessions.}
The router creates a \emph{soft reservation} only after a model call completes.
From the completion metadata, it records the session id, model, serving
instance, reusable-prefix handle, and expiry time
$t_{exp}=t_{done}+\tau$. We choose $\tau=1$\,s from the sensitivity sweep in
\S\ref{sec:eval-reservation}; it is near the p90 tool gap in our traces. The
reservation is an admission lease, not a
GPU-KV pin or an execution guarantee: while live, it charges one configured
admission unit against the serving instance's next eligible model-slot budget.

When the next call for the same session and model arrives, the router checks
that the lease is live, the instance is healthy, and the prefix remains
device-resident or host-restorable. A successful lookup consumes the lease and
routes the call to that instance. A successor lease is installed only after
this call completes; routing does not refresh the old lease. If any check fails,
the router discards the lease and falls back to normal residency-aware
placement.

The lease reduces the chance that unrelated arrivals consume all predicted
admission capacity during a tool gap, but it cannot guarantee same-slot
admission or a TTFT deadline: the next prompt suffix and concurrent returns are
unknown when the lease is created. KV may also move from device to the host
registry under memory pressure. Reservation therefore preserves bounded
affinity and accounts for likely demand; it does not guarantee device residency.
Longer leases preserve more affinity but hold budget longer and can reduce model
residency hit rate; \S\ref{sec:eval-reservation} measures this tradeoff.

\paragraph{Placement for non-reserved requests.}
Requests without an active reservation---new sessions and sessions whose
reservation has expired---are placed by a cost-ranking policy.
The router scores each candidate instance $i$ by
\begin{align}
\text{Cost}(r,i) = \Delta\widehat{R}^{req}_i(r)
 + \widehat{C}^{KV}_i(r)
 + \widehat{C}^{act}_i(m_r),
\label{eq:router-cost}
\end{align}
where all terms are in latency units. The three terms charge incremental
queue/prefill/decode work, KV recovery, and model activation exactly once.
$\Delta\widehat{R}^{req}_i$
contains only the incremental queue, prefill, and decode work introduced by
$r$. State recovery and model activation are excluded from this term and
charged once by the remaining terms.
$\widehat{C}^{KV}_i$ is zero for a device-resident prefix, the measured H2D
restore cost for a host-restorable prefix, and the profiled recomputation cost
otherwise. $\widehat{C}^{act}_i$ is zero when $m_r$ is active, the advertised
switch cost when its state is resident or staged, and the measured full-open
cost otherwise. Partial staging changes the activation estimate rather than
adding a second model-open penalty.

The router refreshes its per-model latency profiles from recent round summaries
using an exponentially weighted moving average. The in-flight overlay in
\S\ref{sec:impl-router} adds calls and reservations forwarded since the latest
snapshot. For each candidate, the router simulates the resulting round load and
HBM use, retaining it only if both configured admission checks pass. These are
predictive overload controls rather than hard TTFT guarantees under stale
telemetry. Miscalibration affects placement quality, while deadline-risk
telemetry makes an underestimated instance less attractive in subsequent
decisions.

$KV_i(s)$ is valid when instance $i$ reports session $s$'s prefix as either
device-resident or host-restorable through HKVR with an agreed prefix.
A new model replica is opened on instance $j$ only when the predicted
demand benefit exceeds the opening cost:
\begin{align}
\text{gain}(m,j) > T_{swap}(m,j) + \mu_j\Delta W_m + \nu_m ,
\label{eq:replica-open}
\end{align}
where $\text{gain}(m,j)$ is the latency-equivalent benefit of opening the
replica: the predicted reduction in queued or spillover demand for model $m$
over the next control window, computed from the router's observed arrivals and
reserved returns. $T_{swap}$ is the switch latency, $\Delta W_m$ is the added HBM
footprint, $\mu_j$ is the instance-specific HBM pressure price, and $\nu_m$ is
a per-replica penalty against unjustified expansion.
$\mu_j$ has units of latency per GiB and converts added HBM footprint into a
latency-equivalent cost using instance $j$'s current memory pressure and staging
slack.

\begin{algorithm}[t]
\caption{Session-aware residency routing.}
\label{alg:router}
\begin{algorithmic}[1]
\REQUIRE Request $r$ for session $s_r$, model $m_r$; instance snapshots and
gateway queue/reservation overlay
\STATE \textit{// Consume a lease created by the previous completion}
\IF{$s_r$ has live lease $(s_r,m_r,h,k,t_{exp})$}
  \IF{$h$ is healthy \AND $KV_h(s_r)$ is valid}
    \STATE consume lease; route $r$ to $h$
    \STATE \textbf{return}
  \ELSE
    \STATE discard lease
  \ENDIF
\ENDIF
\STATE \textit{// Cost-based placement}
\STATE $\mathcal{C} \gets \emptyset$
\FORALL{instances $i$}
  \STATE simulate request, KV, activation, round, and HBM costs
  \IF{round-admission and HBM checks pass}
    \IF{$i$ needs no fresh replica \OR Eq.~\eqref{eq:replica-open} passes}
      \STATE $\mathcal{C} \gets \mathcal{C} \cup \{i\}$
    \ENDIF
  \ENDIF
\ENDFOR
\IF{$\mathcal{C}=\emptyset$}
  \STATE invoke the configured gateway overload policy
  \STATE \textbf{return}
\ENDIF
\STATE $i^* \gets \arg\min_{i\in\mathcal{C}}\,\text{Cost}(r,i)$
\STATE route $r$ to $i^*$
\STATE \textbf{return}
\end{algorithmic}
\end{algorithm}

Algorithm~\ref{alg:router} handles call arrival only. The completion hook in
\S\ref{sec:impl-router} installs a lease for a possible subsequent call. The
lease accounts for a likely return and preserves instance affinity; SP provides
a budgeted opportunity to consume that return while the model slot remains
open. Neither mechanism alone guarantees same-slot admission.
Within the configured cold pool, model-residency preference concentrates demand
on useful instances, while the feasibility filter redirects overflow to nodes
with sufficient round budget and staging slack (remaining capacity for D2D
pre-staging, defined in \S\ref{sec:substrate}).

\subsection{Cold-Pool Scheduler with Session-Prefill (SP) Admission}
\label{sec:scheduler}

\begin{figure}[t]
  \centering
  \includegraphics[width=\columnwidth]{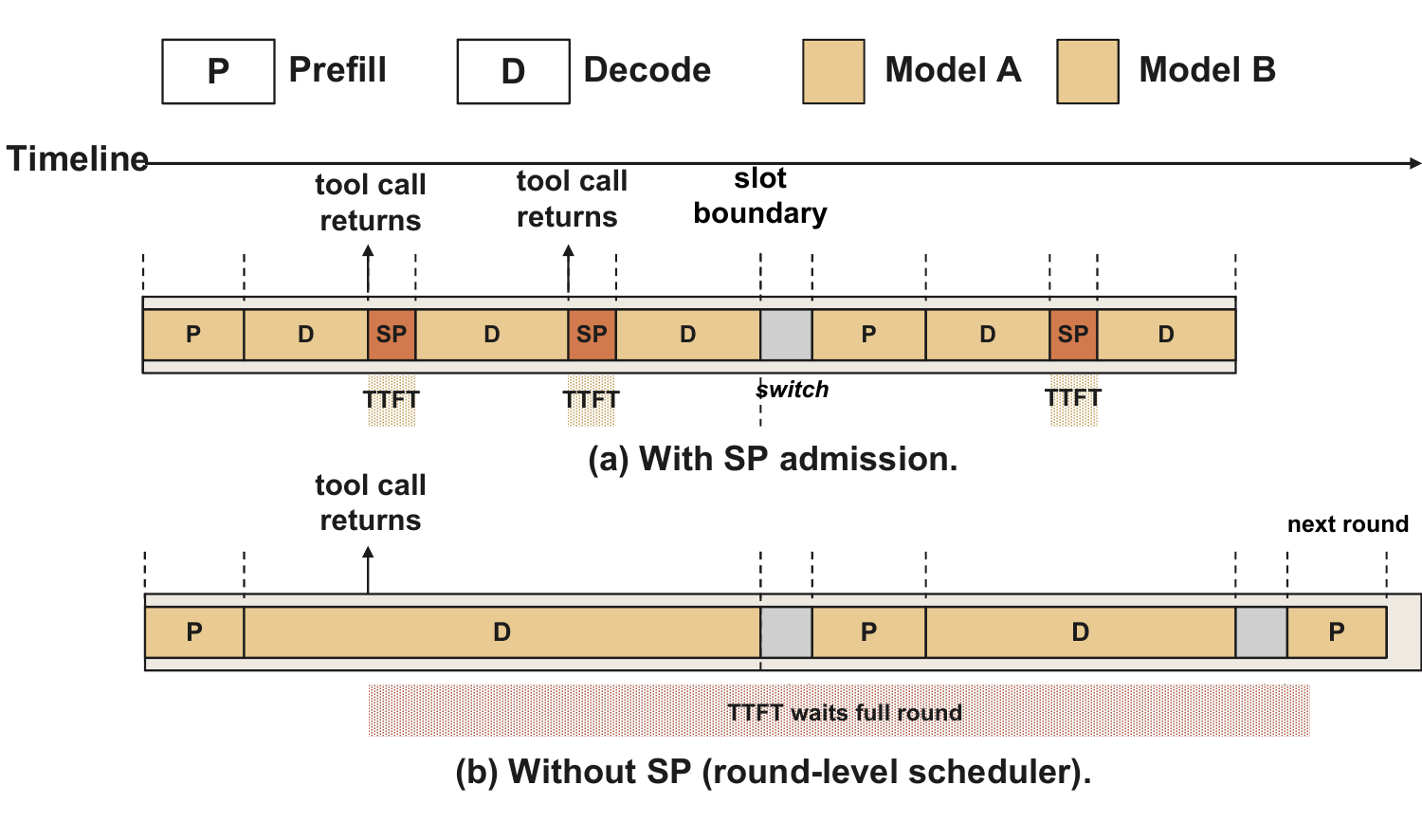}
  \caption{Cold-pool slot structure. SP opens mid-slot prefill windows for
  eligible continuations before the scheduler rotates away.}
  \label{fig:sp-timeline}
\end{figure}

Cold-pool instances must share a GPU group across multiple models while still
serving agent sessions with short tool gaps.
The evaluated design co-locates prefill and decode on the same GPU group,
avoiding a cross-tier KV handoff on each turn. A round-based multi-model
scheduler then faces a second problem:
a session returning from a tool call after its model's slot has closed must
wait until the next full rotation even when its prefix remains reusable.
The SP stage resolves both: by staying non-P/D-disaggregated it avoids
per-turn KV handoff, and by opening a mid-slot admission window it captures
eligible returns before the active model slot closes.

\systemname{} implements a non-P/D-disaggregated token-level scheduler for
the cold pool. A cold-pool instance serves several models on the same
tensor-parallel group, executing only one model at any instant. Time is
organized into rounds; each active model receives one slot per round, and the
instance switches models only at slot boundaries. A slot begins with a
queued-prefill phase $P$, which admits ordinary work already waiting when the
slot opens, and an initial decode phase $D_1$. Zero or more
session-prefill/decode pairs may follow before a final decode phase:
\[
P \;\rightarrow\; D_1 \;\rightarrow\; \bigl(SP \rightarrow D\bigr)^{*} \;\rightarrow\; D_{\text{fin}} .
\]
An SP opportunity occurs only at a decode boundary; it does not preempt an
executing kernel. A call is SP-eligible when it targets the active model, is a
continuation that arrived after $P$, and has a device-resident or
host-restorable prefix on this instance. At the $k$-th opportunity, the
scheduler forms the eligible set $\mathcal{E}_{m,k}$ and selects a bounded batch
$\mathcal{B}_{m,k}\subseteq\mathcal{E}_{m,k}$ whose predicted prefill, restore,
and HBM costs fit the budget remaining after protected decode. Calls that do
not fit remain queued for a later SP opportunity or the next model slot. After
the batch completes, decode resumes for all in-flight requests.
The final decode phase $D_{\text{fin}}$ is the drain interval after the last SP
opportunity in the slot: the scheduler stops admitting new prefill but spends
the remaining protected budget on in-flight decode before switching models.
New sessions do not bypass $P$ through SP. The slot closes when the round
budget is exhausted or the scheduler decides to switch models. A round on
instance $i$ is therefore
\begin{align}
R_i &=
\sum_{m\in M_i^{\text{active}}}
\!\Bigl(P_m + D_{1,m} + \textstyle\sum_{k}(SP_{m,k} + D_{m,k}) + D_{m,\text{fin}}\Bigr)
\nonumber\\
&\quad+ \sum_{(m\rightarrow m')} S_{m\rightarrow m'} ,
\end{align}
where
$SP_{m,k}=\widehat{T}_{pf}(\mathcal{B}_{m,k})+
\widehat{T}_{restore}(\mathcal{B}_{m,k})$ is the batched prefill and restore
cost at the $k$-th opportunity, and $S_{m\rightarrow m'}$ is the switch cost
exposed by the memory substrate (\S\ref{sec:substrate}). The switch estimate is
updated from staging coverage and dirty-chunk telemetry rather than treated as
a static model constant.

\paragraph{Decode-share budget.}
The round scheduler controls average token cadence; it does not claim a hard
deadline for every inter-token gap. Let $d_m$ be the number of decode
iterations reserved for model $m$ in a predicted round, and let $t_m^D$ be its
profiled decode-step time. Define
\[
U_m = R_i - d_m t_m^D
\]
as all predicted time in the round not spent on $m$'s decode iterations,
including its own $P/SP$ phases, other model slots, and switches. For a sequence
that remains active throughout the round, the predicted average time per token
is $R_i/d_m$. To keep this average below
$\overline{T}_{\text{TPOT}}$, the scheduler requires
\begin{align}
d_m\bigl(\overline{T}_{\text{TPOT}}-t_m^D\bigr) \ge U_m .
\label{eq:decode-budget}
\end{align}
If $\overline{T}_{\text{TPOT}}\le t_m^D$, the target is infeasible for that
model profile. The scheduler reserves the corresponding $d_m$ decode
iterations before assigning the remaining slot budget to $P$ and SP.
Intuitively, protected decode establishes the target average token cadence;
prefill and SP share only the time left after that reservation.
The round itself is bounded by the TTFT budget with a safety margin:
\begin{align}
R_i \le \gamma T_{\text{TTFT}}, \quad \gamma < 1 .
\end{align}
This round-length constraint bounds predicted model revisit time; it is not by
itself a hard TTFT guarantee because queueing across rounds and estimation error
can still violate the target.

Prefill arrivals and prompt lengths are unknown when a round begins, so
prefill is treated as best-effort within the remaining slot budget.
The key departure from a standard round-based scheduler is the SP step:
whenever an eligible continuation returns during a slot, it is queued for the
next SP opportunity rather than immediately deferred until the next slot for
that model. At that opportunity, the scheduler admits only the bounded subset
that fits the remaining budget; it never delays protected decode to admit an
unbounded return batch. Decode then resumes for all in-flight sessions,
including newly admitted continuations.
This interleaving can repeat multiple times within a slot as successive tool
calls complete, preserving useful locality and avoiding an unnecessary
next-slot wait.
A continuation that arrives after the slot closes, or does not fit before the
final decode phase, waits for the next slot of the same model.

Best-effort prefill does not mean silently deferring work.
When the protected decode budget leaves insufficient prefill time, or when
cumulative pressure pushes the round past the TTFT bound, the instance
reports the shortfall through structured telemetry---\emph{staging slack}
(remaining capacity for D2D pre-staging, detailed in \S\ref{sec:substrate})
and deadline-risk counters---rather than hiding overload by rolling work into
later rounds.
The router reacts by spilling traffic to another instance, opening a replica,
or applying the configured admission-control policy.

\subsection{Multi-Model Memory Substrate}
\label{sec:substrate}

The router and cold-pool scheduler decide \emph{where} and \emph{when} model
weights and session KV should reside. They rely on a memory substrate that makes
these placement decisions executable within the round budget. The substrate
exposes three capabilities to the rest of \systemname{}.

\textbf{Stable HBM layout.}
Cold-pool instances switch among models whose weights, KV pools, runtime
buffers, and graph state all compete for HBM. The substrate gives each instance
a stable multi-model layout: structural regions can be released, reused, and
rebound across switches without fragmenting device memory or invalidating the
runtime state needed to serve the next slot.

\textbf{Persistent session KV.}
Switching away from a model may reclaim its device KV region, but a returning
session still needs the prefix it produced before the tool call. The substrate
therefore maintains a host-side KV registry that stores valid session prefixes
across model switches and restores them to device memory when the session
returns. The registry is part of the correctness boundary: it must not expose a
prefix unless all tensor-parallel ranks agree on the same valid blocks.

\textbf{Reduced switch cost.}
Model switching consumes the same round budget that SP admission depends on.
The substrate stages future model weights while the current model is serving,
then uses device-to-device movement for the staged portion at switch time and
falls back to host-to-device loading only for uncovered chunks. The router sees
the remaining staging capacity as \emph{staging slack}; when that slack is low,
the instance becomes a worse target for additional traffic. Staging slack is
therefore both a memory signal and a switch-cost signal: lower slack predicts a
larger uncovered H2D fraction on the next switch.

These capabilities form the contract between scheduling and implementation:
the instance reports measured restore and switch estimates, and the router and
SP scheduler charge those estimates to their admission budgets. Section
\ref{sec:implementation} describes how \systemname{} implements the lease and
memory substrate.

%% file: sections/04-implementation.tex
\section{Implementation}
\label{sec:implementation}

We implement \systemname{} on SGLang~\cite{zheng2024sglang} as a stateful Go
gateway (5.5K lines) and an extended inference engine (15K lines) that adds
session caching, HKVR, SP scheduling, and staged switching. We additionally
adapt 1.4K lines of SGLang's existing TMS support for per-model memory
reclamation. The gateway maintains session leases and routing telemetry; each
engine instance implements the instance-local cache, scheduling, and memory
contracts.

\subsection{Router State and Session Cache Integration}
\label{sec:impl-router}

The gateway extracts a stable session id from the agent runtime and keeps a
session table keyed by that id. Each entry records the last serving instance,
model, observed KV length, last-use time, and a soft-reservation lease. After a
request finishes, the gateway updates the entry from response usage metadata and
installs a lease on the serving instance. Before routing the next call, it
overlays its local in-flight queue on top of the latest instance snapshot, so
recent forwards and reservations affect admission immediately instead of waiting
for the next telemetry poll. Instance snapshots are versioned and refreshed on
the control-plane heartbeat; if a snapshot is stale or the instance misses a
heartbeat, the gateway can still honor live reservations only when the instance
passes the health check, but it excludes that instance from new non-reserved
placements until fresh telemetry arrives.

The lease is represented as \((s,m,i,k,t_{exp})\): session \(s\), model \(m\),
reserved instance \(i\), prefix handle \(k\), and expiry time \(t_{exp}\). A
live lease contributes one configured admission-unit charge to the gateway's
local load overlay. On a matching return, the gateway consumes the lease before
forwarding the call to \(i\); the completion hook may later install a successor
lease. If the lease expires, the instance becomes unhealthy, or the prefix is no
longer restorable, the call falls back to normal placement. Because the next
suffix length is unknown at reservation time, this charge is capacity
accounting rather than a hard latency guarantee.

Each SGLang instance makes the prefix handle executable through a
session-aware prefix-cache wrapper. When a turn finishes, the wrapper saves the
request-pool slot, committed KV length, radix lock state, and any
architecture-specific cache state. On the next turn, it restores that slot
before prefix matching, so the existing scheduler sees a prefix hit without a
separate token path. The same finish path checkpoints aligned KV blocks to HKVR
when host KV is enabled. Instance telemetry reports the fields the gateway
needs for routing: loaded models, cached sessions, model queues, round
remaining time, deadline risk, KV occupancy, and free staging bytes.

\subsection{Bump-Managed HBM Layout}
\label{sec:impl-bump}

A cold-pool instance switches among models whose weights, KV tensors, CUDA
graphs, runtime workspaces, and bookkeeping state all occupy HBM. Rebuilding
these objects on every activation would recapture CUDA graphs, reinitialize
attention backends, rebuild KV-pool shells, and fragment PyTorch's caching
allocator. \systemname{} therefore removes the large structural regions from
the caching allocator and places them in one managed HBM buffer. For
PyTorch-owned per-model objects, it uses SGLang's existing
\texttt{torch\_memory\_saver} (TMS) integration, a VMM-based physical-page
manager that preserves virtual addresses while reclaiming physical pages.
\systemname{} adapts that mechanism with per-model tags and switch-time
pause/resume calls, so CUDA-graph and runtime allocations can be parked with
the model that owns them. TMS is a memory mechanism below the scheduler, not
the scheduling abstraction itself.

\begin{figure}[t]
  \centering
  \includegraphics[width=\columnwidth]{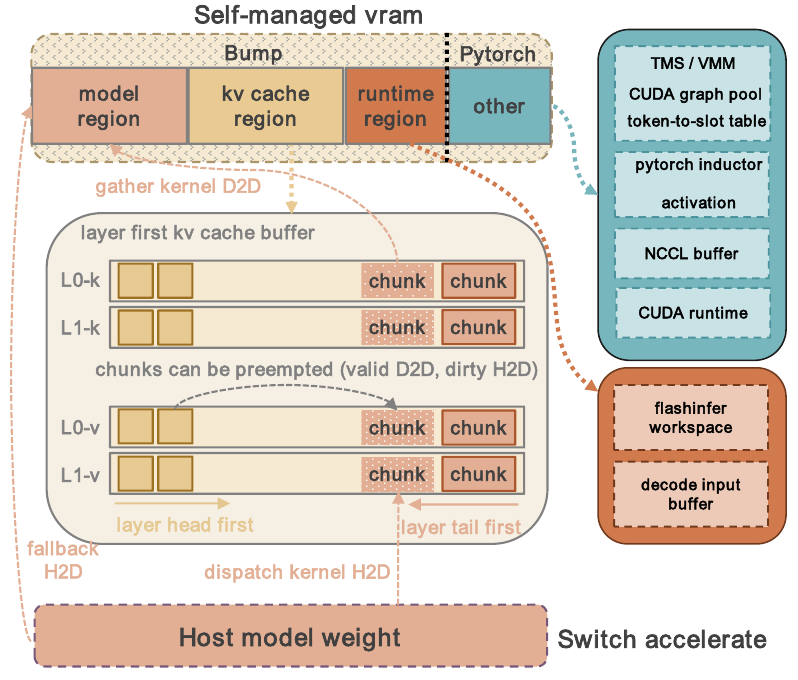}
  \caption{Bump-managed HBM layout. Stable regions preserve pointers across
  switches; clean staged chunks move by D2D and dirty chunks reload from host.}
  \label{fig:bump-allocator}
\end{figure}

The bump buffer uses a fixed layout: weights and KV data grow upward from the
low-address end, runtime buffers are anchored at the high-address end, and the
gap between them forms the KV pool. This layout lets buffer-like regions be
shared across models without static per-model partitions. TMS covers the
PyTorch-managed objects that cannot simply be rebound into the bump buffer,
including CUDA-graph pools and per-model runtime metadata. When a model is
parked, TMS releases their physical pages while preserving virtual addresses;
on restore, the same addresses are remapped, so captured kernel pointers remain
valid.

Switching from model \(m\) to \(m'\) then reduces to releasing the active
\texttt{kv\_cache} region, resizing the \texttt{weights} region for \(m'\), and
rebinding \(m'\)'s parameter tensors to views in the bump buffer. The engine
caches each model's CUDA graph, attention backend, and KV-pool shell, and
recaptures only if a restore-time pointer check detects an address change. The
check compares the base addresses of the weight, KV, graph-pool, and runtime
buffer regions. Once the bump layout is stable, the common switch path reuses
captured kernels directly and avoids \texttt{cudaFree}/\texttt{cudaMalloc}
churn. The stable addresses keep cached CUDA graphs and runtime objects valid
across model switches.

\subsection{Host KV Registry}
\label{sec:impl-hkvr}

Switching away from a model may reclaim its device KV region, but returning
sessions still need the prefix produced before the tool call. HKVR stores this
state in a unified pinned-memory pool shared across attention layouts. MHA
models allocate paired K/V slabs, MLA models allocate latent-state slabs, and
Mamba-style models allocate recurrent-state slabs. All slabs return to the same
pool when evicted, so host memory is not statically partitioned by model or
architecture.

HKVR checkpoints aligned KV blocks asynchronously when a request finishes or a
session enters a tool gap. Each block is keyed by the model name and cumulative
prefix hash, matching the radix tree's chunk granularity. On return, the
scheduler allocates new device slots, restores the agreed prefix through H2D
layer-first/page-first transfer kernels, and recomputes only the unconfirmed
tail. This path reuses SGLang's KV movement kernels for MHA. MLA and Mamba share
the registry metadata and slab allocator, while architecture-specific
completeness checks may decline a restore rather than publish partial state.
The router sees a common device/host/missing state; the instance decides whether
the architecture-specific state is restorable.

Three invariants keep HKVR safe under concurrent serving. First, transfer pins
are separated from radix ownership: DMA pins prevent reclaiming blocks in use,
while radix references track whether prefix metadata still points to a block.
Second, HKVR exposes a prefix only after TP ranks agree on the same valid common
prefix. A disagreement is handled conservatively: the restore tracker truncates
the published prefix to the longest common block range shared by all ranks and
marks the inconsistent tail for recomputation. Partial or inconsistent tails
remain private until a later checkpoint extends the agreement. Third, every
model switch drains outstanding D2H events
before freeing the active \texttt{kv\_cache} region. Since D2H is issued as
turns finish rather than batched at switch time, this barrier usually waits only
for the final tail blocks.

\subsection{D2D-Staged Switching}
\label{sec:impl-switch}

Reloading a multi-hundred-GB model from host at the switch boundary would spend
the same round budget that SP relies on. \systemname{} instead pre-stages the
next model's weights into unused HBM while the current model is serving, then
uses D2D movement for the staged fraction at switch time. The staging surface is
the high-address tail of each KV layer block; the KV allocator reuses low
addresses first, keeping those tails clean under normal allocation pressure.

Staging is opportunistic and correctness-preserving. The staging manager writes
weights into fixed-size chunks and marks a chunk dirty if a concurrent prefill
allocation takes rows that overlap the chunk's staging range. During an active
H2D write, the allocator skips the chunk's row range to avoid reading partially
written data. At switch time, the gather plan copies only clean staged chunks
with D2D and reloads dirty or missing chunks from host.

This protocol makes the expensive H2D component scale with the uncovered weight
fraction; the total switch still includes D2D gathering for staged chunks. It
also gives the router a simple signal:
\emph{staging slack}, the remaining HBM space that can be used to pre-stage a
future switch. When slack falls, the instance becomes a worse target for more
traffic in the residency cost model; Section~\ref{sec:eval-substrate} separates
the full-D2D, partial, and H2D fallback paths.
Together, the router lease, bump layout, HKVR, and D2D staging realize the
substrate contract exposed to the design: stable residency, restorable session
state, and measured switch-cost telemetry.

%% file: sections/05-evaluation.tex
\section{Evaluation}
\label{sec:eval}

We evaluate \systemname{} with fixed replay as the primary evidence and
targeted mechanism measurements for effects that a single replay cannot
separate. Fixed replay holds the application-level call trace constant, so
policy differences are not confounded by the agent choosing a different tool
path. We ask five questions: (1)~do the mechanisms improve
end-to-end session completion, (2)~does residency-aware routing avoid misplaced
model opens, (3)~what is the cost/benefit of soft reservation, (4)~does SP
capture returns within an active slot, and (5)~does the memory substrate reduce
switch cost without exceeding HBM capacity?

\subsection{Methodology}
\label{sec:eval-setup}

The main replay is captured from Claude Code running ten Astropy issues from
SWE-Bench Verified~\cite{jimenez2024swebench}. We replay each issue once on
Qwen3-235B, GLM5-nvfp4, and Qwen3.5-122B-A10B, yielding 30 model-session
executions and 960 calls. Reusing the ten task identities across three model
profiles enables paired, within-trace policy comparisons; the per-session plot
shows every observation.

For each model-session, every configuration replays the same request bodies
under the same model and session IDs, in the same per-session call order, and
with the same completion-token count for each call. The replay is closed-loop:
after a call finishes, the next waits for the same recorded tool gap after a
10\,s cap. The cap affects 20 of 960 calls. Global interleaving may differ across
configurations because completion times differ. The resulting trace targets the
long-context, short-return-gap regime for which session continuity matters most.
Large-model runs use one TP=8 server with eight 183\,GiB GPUs, 2\,TB host
memory, dual-socket x86 CPUs, and eight 400\,Gb/s network adapters. Each run
performs the same pre-window warmup; all measurements exclude warmup.

We compare \emph{Talaria}'s instance-local configuration (SP, HKVR, and D2D
staging) with reverse ablations. \emph{no-SP} disables mid-slot
session-prefill admission while keeping HKVR and D2D staging. \emph{no-HKVR}
keeps SP and D2D staging but disables host-restorable KV. \emph{H2D-only} keeps
SP and HKVR but disables D2D weight staging, so switches reload from host
memory. \emph{Round-only} disables SP, HKVR, and D2D staging and serves the
same trace with the otherwise identical round scheduler and host-to-device
model reloads. Round-only is the controlled all-off baseline within the same
engine.

Router-policy experiments use two TP=4 workers to create real placement
choices. They replay a 120-call, 30-model-session subset under three relative
offered-load regimes. The trace preserves session ids, return gaps, model order,
and reusable-prefix lengths, but maps the large models to Qwen3-32B,
Qwen2.5-32B-Instruct, and Qwen3.5-35B-A3B. This split is intentional: TP=8
evaluates instance-local mechanisms under large-model memory pressure, while
TP=4 isolates placement decisions. Together, the two testbeds separate
instance execution from cluster placement without conflating their effects.

The primary metric is \emph{session completion time} (SCT), measured from a
model-session's first call to its final response. We also report TTFT, switch
latency, round composition, and HBM footprint. TTFT is schedule-aware: it starts
at the intended enqueue time and therefore includes admission delay. For raw
replay distributions, percentiles use the nearest-rank empirical definition:
for $N$ observations, $p_q$ is the observation at rank
$\lceil qN/100 \rceil$; targeted TP=4 experiments retain their original
harness summaries.
The main trace intentionally runs at high pressure to expose differences in how
the policies preserve session state while per-session call order, bodies, and
recorded gaps remain fixed.

\subsection{End-to-End Session Completion}
\label{sec:eval-e2e}

\begin{figure}[t]
\centering
\includegraphics[width=\columnwidth]{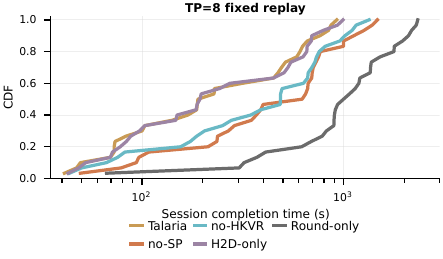}
\caption{End-to-end SCT on the TP=8 fixed replay. Talaria cuts p50 SCT from
1000\,s to 189\,s relative to the Round-only ablation.}
\label{fig:eval-sct-cdf}
\end{figure}

\begin{figure}[t]
\centering
\includegraphics[width=\columnwidth]{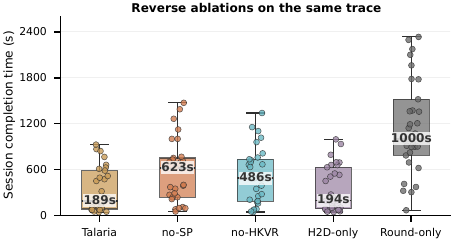}
\caption{SCT for all 30 model-sessions on the same replay; boxes show the
interquartile range under the nearest-rank empirical definition. Each point is
one of ten issues executed on one model.}
\label{fig:eval-sct-ablation}
\end{figure}

Figures~\ref{fig:eval-sct-cdf} and~\ref{fig:eval-sct-ablation} show the SCT
distribution for the TP=8 fixed-replay configurations. All rows replay the
same 960 calls over the same 30 sessions.

Relative to Round-only, Talaria's instance-local configuration reduces p50 SCT
from 1000\,s to 189\,s and p95 SCT from 2296\,s to 867\,s, delivering
5.3$\times$ and 2.6$\times$ speedups. The p50 TTFT falls from 13.44\,s to
0.55\,s. The reverse ablations
show each mechanism's marginal effect with the others enabled; these effects
are not additive. Round-only combines next-slot waiting with prefix
recomputation after reusable KV is lost. SP provides the largest p50 marginal
contribution, reducing SCT from 623\,s to 189\,s by admitting eligible returns
within the active model slot. HKVR reduces p50 SCT from 486\,s to 189\,s and
TTFT p95 from 26.33\,s to 14.17\,s. D2D staging leaves p50 SCT nearly unchanged
(194\,s to 189\,s) but reduces p95 from 933\,s to 867\,s;
\S\ref{sec:eval-substrate} isolates its switch path. Talaria improves 29 of 30
paired model-sessions.

\subsection{Spatial Placement: Router Policy}
\label{sec:eval-router}

\begin{figure}[t]
\centering
\includegraphics[width=\columnwidth]{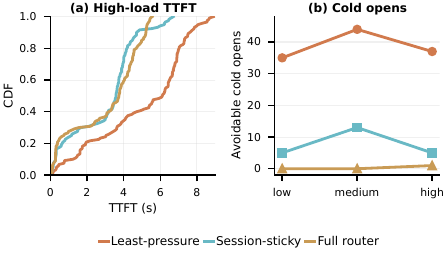}
\caption{Two-worker TP=4 placement over 120 calls. The full router nearly
eliminates avoidable model opens while controlling high-load TTFT.}
\label{fig:eval-router-placement}
\end{figure}

Figure~\ref{fig:eval-router-placement} compares three routing stacks on the same
120-call trace.
\emph{Least-pressure} chooses the least-loaded worker for each call.
\emph{Session-sticky} pins a session to its first worker. The \emph{full
router} combines residency cost ranking with soft reservations, matching the
prototype policy; \S\ref{sec:eval-reservation} separately sweeps the reservation
timeout. We count an avoidable cold open when the selected worker must open the
requested model even though another worker already has that model resident.

Least-pressure frequently selects a worker that must reopen the requested
model. Session-sticky preserves affinity but cannot trade it against model
residency and queue pressure. The full router eliminates avoidable opens at low
and medium load and leaves one at high load. At medium load, TTFT p50 is
0.36\,s, versus 1.67\,s for least-pressure and 2.29\,s for session-sticky. At
high load, it retains 97.8\% of returning calls on the same worker, reduces
avoidable opens from 37 (least-pressure) and 5 (session-sticky) to 1, and cuts
TTFT p95 from 8.07\,s and 6.23\,s to 5.28\,s.

The placement gains translate most clearly once model opens and queue pressure
matter. At medium load, full routing improves SCT p50/p95 to 6.28/12.74\,s
from 11.01/19.59\,s for least-pressure and 10.97/19.28\,s for session-sticky.
At high load, it lowers p95 to 21.31\,s from 31.29\,s for least-pressure and
23.03\,s for session-sticky; its 17.44\,s p50 is slightly above sticky's
16.33\,s. At low load, eliminating avoidable opens does not translate into an
SCT gain: full-router p50/p95 is 7.39/15.81\,s, versus 6.82/11.31\,s for
least-pressure and 5.62/13.68\,s for session-sticky. Thus residency-aware
placement yields its clearest SCT benefit at medium load; at high load it
improves p95 but not p50, while at low load it eliminates model opens without
an SCT gain.

\subsection{Soft-Reservation Tradeoff}
\label{sec:eval-reservation}

Soft reservation adds a temporal decision to router placement: after a session
leaves for a tool call, the gateway leases capacity for a bounded interval
$\tau$ instead of making the next call compete as an unrelated arrival. The
TP=4 sensitivity run sweeps $\tau \in \{0,0.5,1,2,5\}$\,s and reports
same-instance return rate, model-residency hit rate, returning-call TTFT, model
opens, and an unused-lease proxy. The proxy counts reserved returns that
did not produce a timely same-instance hit; it does not measure how long a live
lease occupied capacity.

\begin{figure}[t]
\centering
\includegraphics[width=\columnwidth]{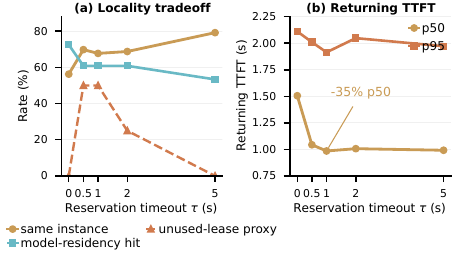}
\caption{TP=4 reservation-timeout sweep. A 1\,s lease gives the lowest
returning-call p95; extending to 5\,s trades model residency for higher return
locality.}
\label{fig:eval-reservation}
\end{figure}

With $\tau=0$, returning sessions still use residency-aware cost ranking, but
no capacity is held: 56.2\% return to the same worker and 43.8\% miss the
device-KV proxy. A 1\,s lease raises same-worker returns to 67.7\% and lowers
returning-call TTFT p50/p95 from 1.51/2.11\,s to 0.99/1.91\,s, the lowest p95
in the sweep. This gain uses additional placement capacity: relative to
$\tau=0$, model-residency hits fall from 72.5\% to 60.8\%, model opens rise
from 33 to 47, and the unused-lease proxy is 50\%. Extending the lease to 5\,s
raises same-worker returns further to 79.2\%, but provides similar TTFT while
model-residency hits fall to 53.3\% and model opens rise to 56. We therefore
use 1\,s as the operating point that best balances return locality and model
residency for this workload; $\tau$ is a workload-dependent policy knob.

\subsection{Temporal Placement: SP Admission}
\label{sec:eval-sp}

\begin{figure}[t]
\centering
\includegraphics[width=\columnwidth]{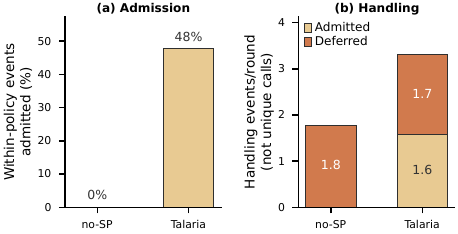}
\caption{SP handling events on the TP=8 fixed replay. Talaria admits 47.8\% of
its recorded return-handling events within the active model slot.}
\label{fig:eval-sp-admission}
\end{figure}

\begin{figure}[t]
\centering
\includegraphics[width=\columnwidth]{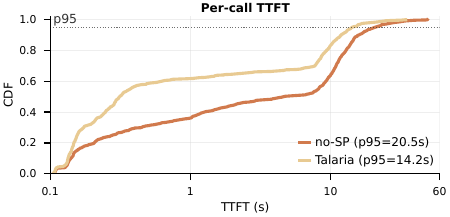}
\caption{SP TTFT impact on the same replay. Mid-slot admission lowers per-call
TTFT.}
\label{fig:eval-sp-ttft}
\end{figure}

SP addresses the temporal mismatch inside a model slot: a returning session
should not wait for the next full prefill window when the current slot can
admit its reusable prefix. Figures~\ref{fig:eval-sp-admission}
and~\ref{fig:eval-sp-ttft} compare Talaria with no-SP on the same fixed replay,
keeping request bodies, arrival gaps, HKVR, D2D staging, and the router fixed.

With all other mechanisms fixed, SP lowers p50 TTFT from 4.60\,s to
0.55\,s and p95 from 20.5\,s to 14.2\,s. Within Talaria's 200 rounds, SP admits
316 return-handling events into the active slot and parks 345, an admission
rate of 47.8\%. Talaria records none of the 123 host-cache capacity misses and
2{,}532 architecture-specific restore skips observed with no-SP. These are
schedule-dependent diagnostics, not unique-call outcomes; all 960 calls
complete in both configurations. A parked request may also appear in multiple
rounds, so we normalize handling events within each policy and use TTFT,
measured over the common call set, as the direct cross-policy result.

\subsection{Memory Substrate: Switch and HBM Cost}
\label{sec:eval-substrate}

\begin{figure}[t]
\centering
\includegraphics[width=\columnwidth]{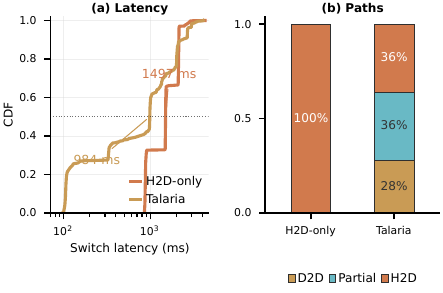}
\caption{Logical TP=8 switch latency, computed as the maximum across ranks.
Each configuration contains 600 logical switches; paths use the worst-rank
classification. Staging cuts p50 by 34\%.}
\label{fig:eval-switch-path}
\end{figure}

Across the replay, D2D staging reduces aggregate switch time by 38\% and switch
share from 51.7\% to 34.6\%. Figure~\ref{fig:eval-switch-path} isolates the
switch path by comparing Talaria with H2D-only, which keeps SP and HKVR enabled
but reloads weights from host memory.

Each logical switch emits one record per TP rank; we report the maximum rank
latency because the scheduler resumes only after all ranks finish. At this
granularity, D2D staging improves the center of the distribution but exposes a
tail tradeoff under high KV pressure: p50 falls from 1497\,ms to 984\,ms and
mean from 1521\,ms to 1071\,ms, while p95 rises from 2133\,ms to 2679\,ms.
Full D2D switches,
27.7\% of the observed mix, complete in 106\,ms at p50; partial D2D plus H2D
accounts for 36.3\% at a 1000\,ms p50, and the remaining 36.0\% falls back to
H2D at a 2011\,ms p50. The staged common path therefore delivers the
largest speedup, while partial coverage retains a proportional benefit. Active
writes dirty some staged chunks, causing the correctness guard to reload them
from host and accounting for the heavier tail. H2D-only also records 39
host-cache capacity misses and 1{,}704 Mamba restore skips, versus zero for
Talaria. These schedule-dependent diagnostics capture downstream cache
interactions; the logical-switch distribution is the direct path comparison.

\begin{figure}[t]
\centering
\includegraphics[width=\columnwidth]{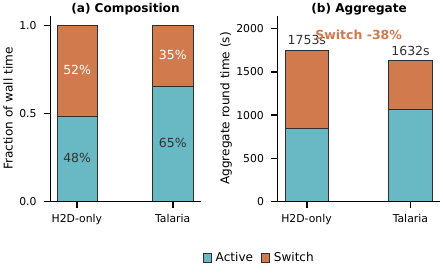}
\caption{Round-time composition for Talaria and H2D-only. D2D staging reduces
aggregate switch time by 38\%.}
\label{fig:eval-round-breakdown}
\end{figure}

Figure~\ref{fig:eval-round-breakdown} shows the same effect at round granularity.
H2D-only spends 51.7\% of round wall time in switching, while Talaria spends
34.6\%; total switch time drops from 907\,s to 565\,s. Talaria's aggregate round
wall time is also lower (1632\,s versus 1753\,s). Because SP and HKVR already
preserve the common return path, the 38\% switch-time reduction has a larger
effect on tail SCT than on p50: p50 changes from 194\,s to 189\,s, while p95
falls from 933\,s to 867\,s.

\begin{figure}[t]
\centering
\includegraphics[width=\columnwidth]{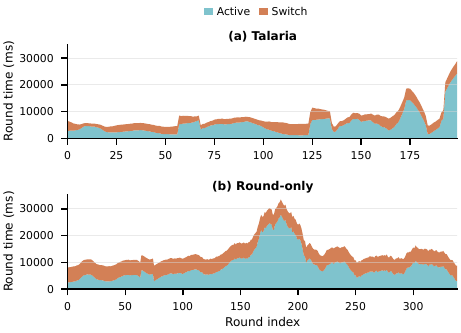}
\caption{Per-round composition for Talaria and Round-only. The combined
mechanisms reduce switch share and round count.}
\label{fig:eval-round-timeline}
\end{figure}

Figure~\ref{fig:eval-round-timeline} shows the combined effect over time.
Round-only carries a persistent switch band because every slot-boundary model
change reloads from host memory. Talaria's SP, HKVR, and staging mechanisms
jointly reduce switch share and the number of rounds; this comparison does not
attribute the difference to staging alone. Curves use an 11-round moving
average for visual clarity and retain separate round-index axes because the
configurations execute 200 and 339 rounds.

\begin{figure}[t]
\centering
\includegraphics[width=\columnwidth]{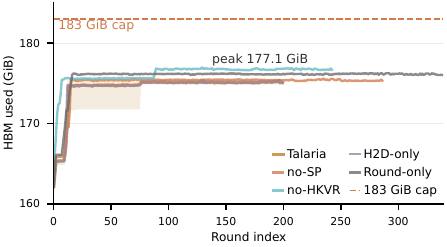}
\caption{HBM footprint under the bump allocator. Runs remain below the
183\,GiB device cap. Curves show per-round averages; the annotation is the
maximum across all configurations.}
\label{fig:eval-memory}
\end{figure}

Figure~\ref{fig:eval-memory} shows device footprint across configurations. The
peak observed HBM use is 177.1\,GiB, below the 183\,GiB device cap. This
confirms that the managed layout, restorable KV, and weight staging remain
within the device-memory budget throughout the replay.

%% file: sections/06-related-work.tex
\section{Related Work}
\label{sec:related-work}

\newcommand{\relatedpara}[1]{\smallskip\noindent\textbf{#1.}\ }

Table~\ref{tab:related-compare} distinguishes KV-aware placement (``KV
route'') from internal KV synchronization: only the former lets residency
change which instance receives a return.

\begin{table}[t]
\centering
\scriptsize
\setlength{\tabcolsep}{2.0pt}
\begin{tabular}{@{}lcccc@{}}
\toprule
System & Unit & Model route & KV route & Agent return \\
\midrule
Aegaeon~\cite{xiang2025aegaeon} & request/token & Yes & No & next model slot \\
Preble~\cite{srivatsa2024preble} & prompt & No & Yes & request arrival \\
Mooncake~\cite{qin2025mooncake} & KV object & No & Yes & storage restore \\
ServerlessLLM~\cite{fu2024serverlessllm} & request & Yes & No & cold start \\
\systemname{} & session & Yes & Yes & reservation + SP \\
\bottomrule
\end{tabular}
\caption{Placement signals and cross-call return handling.}
\label{tab:related-compare}
\end{table}

\relatedpara{Serverless and multi-model LLM serving}
Classical prediction-serving systems focus on model lookup, batching,
replication, pipeline provisioning, and interference control under request-level
latency objectives~\cite{olston2017tensorflowserving,crankshaw2017clipper,
shen2019nexus,crankshaw2020inferline,romero2021infaas,
gujarati2020serving}. Serverless systems similarly hide placement behind a
request/function abstraction whose state is short-lived relative to model
execution~\cite{shahrad2020serverless}. LLM serving makes weights and KV cache
persistent serving state. ServerlessLLM
optimizes checkpoint loading and migration for on-demand LLM inference, and
SpotServe adapts serving to preemptible GPU availability~\cite{
fu2024serverlessllm,miao2024spotserve}; AlpaServe and MuxServe multiplex
multi-model or model-parallel inference on shared GPUs~\cite{li2023alpaserve,
duan2024muxserve}; Aegaeon targets long-tail LLM marketplaces with token-level
auto-scaling~\cite{xiang2025aegaeon}. These systems multiplex models and
requests. \systemname{} instead treats the session as the scheduling object: a
router must preserve model and KV residency across tool gaps, and a cold-pool
instance must admit the returning session before the locality expires.

\relatedpara{Adapter and fine-tuned-model serving}
S-LoRA, Punica, and dLoRA make large catalogs of LoRA adapters practical by
sharing a resident base model, paging adapter state, or dynamically migrating
requests and adapters~\cite{sheng2024slora,chen2024punica,wu2024dlora}. These
systems reduce the cost of serving many fine-tuned variants of a base model.
\systemname{} targets the complementary full-model setting: the active model
itself may switch, and a returning session's KV must remain usable across tool
gaps and model residency changes.

\relatedpara{Memory management for LLM serving}
PagedAttention, vAttention, FlexGen, LoongServe, InfiniGen, and Infinite-LLM
improve LLM serving by reducing KV fragmentation, exploiting virtual memory,
offloading cache state, or distributing long-context KV across memory
tiers~\cite{kwon2023vllm,prabhu2024vattention,sheng2023flexgen,
wu2024loongserve,lee2024infinigen,lin2024infinitellm}. These systems optimize
the memory hierarchy for active requests or long contexts. \systemname{} uses
memory management for a different scheduling contract: a model can leave the GPU
without losing a returning session's prefix, and a future model switch can be
prepared while the current model is still serving.

\relatedpara{Execution engines and phase-aware scheduling}
Orca, vLLM, Sarathi-Serve, FastServe, DeepSpeed-FastGen, and NanoFlow improve
utilization of a loaded model through iteration-level scheduling, KV paging,
chunked prefill, preemption, dynamic split-fuse scheduling, and intra-device
overlap~\cite{yu2022orca,kwon2023vllm,agrawal2024sarathi,wu2023fastserve,
holmes2024deepspeedfastgen,zhu2024nanoflow}. Splitwise, DistServe, and
TetriInfer separate
prefill and decode resources, and TokenScale autoscales such deployments using
token velocity~\cite{patel2024splitwise,zhong2024distserve,hu2024tetriinfer,
lai2025tokenscale}.
\systemname{} is complementary: it coordinates model switching, session returns,
and memory pressure across many models. Its evaluated co-located design avoids
adding a cross-tier KV handoff to each session turn.

\relatedpara{KV reuse and cache-aware routing}
CachedAttention and SGLang reuse conversation or structured-program prefixes
through KV hierarchy and RadixAttention~\cite{gao2024cachedattention,
zheng2024sglang}. Hydragen accelerates shared-prefix attention, CacheGen and
CacheBlend reduce the cost of moving or composing reusable KV, and LMCache
provides a KV-cache layer for enterprise-scale serving~\cite{juravsky2024hydragen,
li2024cachegen,yao2024cacheblend,liu2024lmcache}. Mooncake and MemServe expose
KV cache as a disaggregated cluster resource, while Preble, DualMap, and Llumnix
use cache-aware routing or migration to balance locality and
load~\cite{qin2025mooncake,hu2024memserve,srivatsa2024preble,yuan2026dualmap,
wu2024llumnix}. \systemname{} builds on the same principle that KV is
schedulable state, but the routing decision also depends on model residency,
round pressure, and observed session-return timing.

\relatedpara{Serving agentic programs}
ReAct, Toolformer, ReWOO, Reflexion, and AutoGen popularized tool-interleaved,
reflective, and multi-agent language-model execution, and SWE-agent shows that
software-engineering agents repeatedly call models while interacting with a
repository and test environment~\cite{yao2023react,schick2023toolformer,
xu2023rewoo,shinn2023reflexion,wu2023autogen,yang2024sweagent}. SGLang,
Parrot, Pie, Autellix, and Continuum then show that such programs expose
reusable prefixes, control flow, inter-request dataflow, and KV lifetimes that
should be scheduled above the single-request level~\cite{zheng2024sglang,
lin2024parrot,gim2025pie,luo2025autellix,li2025continuum}. \systemname{}
consumes that session-level structure at a complementary layer: placement and
execution across a serverless multi-model pool. Here, models may be hot or
cold, and the router must jointly decide whether to reuse a model replica,
restore session state, or recompute a prefix.

%% file: sections/07-conclusion.tex
\section{Conclusion}
\label{sec:conclusion}

This paper presented \systemname{}, a serverless multi-model serving system
for agentic workloads. Serving an agent session requires tracking both session
state---KV locality and return timing---and instance state---model residency and
pressure. Request-level placement and slot-boundary admission lose continuity
in different ways. \systemname{}
addresses them with soft reservation at the router and session-prefill (SP) at
the cold-pool scheduler. An instance-local substrate preserves restorable KV
and stages weights across model switches.

On the fixed TP=8 replay, the instance-local mechanisms deliver 5.3$\times$
and 2.6$\times$ speedups in p50 and p95 SCT relative to the internal Round-only
ablation, and improve 29 of 30 paired model-sessions. SP lowers p50 TTFT from 4.60\,s to
0.55\,s, while D2D staging cuts p50 logical-switch latency by 34\% and
aggregate switch time by 38\%. On the two-worker placement testbed, the router
eliminates every avoidable model open at low and medium load and leaves only one
at high load.

The design intentionally keeps the router out of the token data path. Weight
and KV transfers are executed by instances, while the router uses telemetry to
decide when those transfers are worth paying. This separation is what lets
\systemname{} combine cluster-level placement with instance-level execution
contracts: the router selects an instance using model residency, KV
availability, and pressure; that instance restores KV and activates or stages
models as needed.

\paragraph{Limitations and future work.}
Our current evidence combines controlled large-model execution on one TP=8
server with isolated placement experiments on two workers. Broader workload
and cluster-scale evaluation are the next empirical steps. On the mechanism
side, workload-adaptive lease and host-cache policies, remote-KV integration,
dynamic hot-cold pool reallocation, and P/D-disaggregated admission are natural
extensions. D2D staging improves
the p50 and mean switch path under high KV pressure; reducing dirty-chunk
fallbacks can extend these gains to p95.

%% file: sections/appendix.tex
\section{Trace and Replay Details}
\label{app:trace}

\paragraph{Session boundaries.}
A session is one agent task. The trace starts at the first model call issued for
that task and ends at the final model response before the task terminates. We
define the \emph{inter-call gap} as the wall-clock interval between a model
response and the next model enqueue. This interval includes tool execution and
agent/runtime orchestration, so it is not a pure tool-execution measurement.
Harness startup and teardown fall outside the session boundary. Across the
6{,}941 possible consecutive-call intervals, 6{,}294 have a complete timestamp
pair; the remaining 647 do not. The characterization plot further excludes five
${>}30$\,s environment-startup intervals, leaving 6{,}289 gaps for the reported
distribution.

\paragraph{Prefix reuse.}
For consecutive calls within a session, prefix reuse is measured after
tokenization as the longest common prefix between the new prompt and the prior
prompt plus generated response that remains in the session context. The
resulting prefix length quantifies the amount of reusable work. Whether a return
actually reuses device KV, restores host KV, or recomputes the prefix also
depends on cache residency, validity, and capacity at return time.

\paragraph{Replay inputs.}
The 445-session characterization trace supports the distributions and
recovery-cost estimates in \S\ref{sec:bg-session}. Each available record contains
the session id, call index, model id, enqueue timestamp, prompt and output
tokens, reusable-prefix tokens, and inter-call gap. The recovery-cost estimates
combine these records with model-specific prefill profiles measured on the
testbed.

The TP=8 fixed replay is a separate 960-call workload. Across configurations,
each model-session uses the same request bodies, model and session ids,
per-session call order, completion-token counts, and recorded inter-call gaps.
The replay caps each gap at 10\,s, affecting 20 of 960 calls. It is closed-loop:
each next call is released after the preceding call completes and its capped gap
elapses, so global interleaving may differ as configurations finish calls at
different times. The TP=4 router experiments use a separate 120-call,
30-model-session subset and map its model ids to the models on that testbed
(\S\ref{sec:eval-setup}).

\paragraph{Timing telemetry.}
Replay logs keep request-local TTFT and schedule-aware TTFT in the same schema.
The latter starts at the replay scheduler's intended enqueue time, so it
includes any replay-side admission delay. Unless stated otherwise, the paper
reports schedule-aware TTFT. Round summaries use a mutually exclusive
decomposition, \texttt{wall\_ms = active\_ms + switch\_ms}. Phase counters such
as prefill, decode, H2D, and D2H are diagnostic subviews inside active slots;
they can overlap and are not summed into wall time.

\section{Correctness Conditions}
\label{app:correctness}

\systemname{} relies on three implementation-level conditions when moving state
across HBM and host memory.

\textbf{Prefix agreement.}
Each tensor-parallel rank reports its committed prefix length and cumulative
prefix hash at page boundaries. HKVR exposes only the longest page-aligned
prefix on which all ranks agree for the same model. If a rank lags or hashes
diverge, HKVR publishes the last agreed boundary and marks the remaining suffix
for recomputation. This makes TP disagreement a reuse loss rather than a
correctness risk.

\textbf{Lifetime separation.}
Radix-tree ownership and DMA transfer pins are tracked separately. A block can
remain pinned until an asynchronous copy finishes even if the radix tree no
longer references it; conversely, radix metadata cannot expose a block whose DMA
copy has not completed.

\textbf{Switch barrier.}
Before a model's active \texttt{kv\_cache} region is released, the scheduler
drains outstanding D2H events for prefixes that may be restored later. The
barrier is scoped to the outgoing model, so unrelated models do not block the
switch path.

\paragraph{Validation protocol.}
We exercise these guards at two levels. Synthetic switch/restore tests use
deterministic prompts whose first generated tokens are known; after a forced
model switch and HKVR restore, the client compares the generated prefix with the
expected sequence and records whether the cache hit came from device or host
state. Large-model stress runs enable the same runtime checks under round
scheduling: TP-prefix disagreement truncates the exposed prefix, stale or
uncommitted DMA blocks are never published to radix metadata, and staged weight
chunks marked dirty are reloaded from host rather than copied through D2D. These
tests exercise the guards; performance claims use the workloads in
\S\ref{sec:eval}.

\section{HKVR Transfer Layouts}
\label{app:hkvr-transfer}

HKVR separates the scheduler-visible cache state from the physical layout used
to move KV bytes. The router only observes whether a prefix is device-resident,
host-restorable, or missing. The restore path below that interface packs and
unpacks KV pages through layout-aware transfer kernels.

\textbf{Layer-first and page-first movement.}
SGLang's KV movement path supports two equivalent views of the same logical
prefix. In a \emph{layer-first} (LF) layout, host slabs are grouped by model
layer, then by page within that layer; this matches the device-side KV cache
layout used by attention kernels and makes a layer's pages contiguous for DMA.
In a \emph{page-first} (PF) layout, the host representation is grouped by
logical prefix page, with the per-layer fragments for that page stored under
one page descriptor. PF is useful for registry operations because the prefix can
be committed, invalidated, or restored at page granularity. HKVR records the
logical prefix as pages and invokes LF/PF transfer kernels to translate between
the registry view and the active device layout. LF/PF are data-movement layouts;
HKVR is the consistency and naming layer above them.

\textbf{Attention-layout metadata.}
Each HKVR entry carries a compact layout descriptor: model id, TP rank, prefix
hash, committed token length, page size, dtype, attention family, and slab
shape. MHA models store paired K/V slabs, while MLA models use latent-state
slabs. Supported Mamba-style paths register recurrent and auxiliary state
through architecture-specific descriptors rather than treating that state as
KV slabs. The router-visible states remain device-resident, host-restorable, or
missing across these families, but transfer and completeness checks are
architecture-specific. HKVR publishes restorable state only when that state is
complete and the agreement rule in \S\ref{app:correctness} holds; otherwise the
unconfirmed suffix is recomputed.

\section{Switch Timeline Instrumentation}
\label{app:switch-timeline}

The implementation records one \texttt{SWITCH\_TIMING} line per TP rank for
every model switch. Each rank's timer starts after pre-switch diagnostic probes.
The instrumented latency of a logical switch is the maximum across its TP ranks;
the phase breakdown below instead reports rank-level records so that per-rank
work remains visible. A switch follows five ordered phases.

\textbf{Phase 1: save and teardown.}
The instance drains outstanding HKVR D2H writes for the outgoing model, cancels
stale H2D restores, saves the CUDA-graph and KV-pool handles, releases runtime
buffers, parks the outgoing radix tree, and pauses the outgoing model's
per-model TMS tags.

\textbf{Phase 2: open target weights.}
The instance waits for any outstanding pre-staging work, verifies the staging
bitmap, updates the active model configuration, resolves the CPU-side model
entry, and opens the target weights either through D2D gather, partial D2D plus
H2D top-up, or full H2D fallback.

\textbf{Phase 3: restore the KV shell.}
The scheduler checks whether the target model's KV-pool shell can be reused
under the current weight-region size, page size, and capacity. A hit rebinds the
cached shell; a miss rebuilds the pool under the target model's TMS tag.

\textbf{Phase 4: restore runtime and graph state.}
The instance restores a cached attention backend and CUDA graph when the KV
shell and graph pointers match; otherwise it rebuilds runtime buffers and
recaptures graphs.

\textbf{Phase 5: publish references.}
The instance propagates the new model, allocator, KV, radix, and telemetry
references back to the scheduler and worker. This phase is intentionally small:
large memory movement has already completed in Phase~2.

\begin{center}
\captionsetup{type=table}
\centering
\scriptsize
\setlength{\tabcolsep}{2.6pt}
\begin{tabular}{@{}lrrrrrr@{}}
\toprule
Phase &
\multicolumn{2}{c}{D2D full} &
\multicolumn{2}{c}{D2D + H2D} &
\multicolumn{2}{c}{H2D-only} \\
& p50 & p95 & p50 & p95 & p50 & p95 \\
\midrule
P1 & 26.2 & 34.7 & 26.9 & 30.9 & 21.9 & 31.7 \\
P2 & 109.6 & 244.8 & 995.8 & 1819.8 & 2570.6 & 3142.2 \\
P3 & 0.5 & 2.0 & 0.5 & 1.2 & 0.8 & 1.9 \\
P4 & 8.6 & 20.3 & 8.3 & 20.0 & 7.7 & 16.4 \\
P5 & 0.2 & 0.3 & 0.2 & 0.3 & 0.2 & 0.3 \\
\midrule
Total & 149.8 & 298.9 & 1041.2 & 1854.6 & 2608.6 & 3170.6 \\
\bottomrule
\end{tabular}
\captionof{table}{Per-rank five-phase switch timing in a separate component
stress run (ms).}
\label{tab:switch-phases}
\end{center}

Table~\ref{tab:switch-phases} reports a separate TP=8 component stress run, not
the 600 logical switches per configuration in Figure~\ref{fig:eval-switch-path}.
D2D-full and D2D+H2D use steady-state rank-level records after excluding
warm-up and instrumentation outliers (\(n=14{,}952\) and \(n=102\)); H2D-only
uses rank-level records from a no-D2D stress run (\(n=2{,}104\)). The table
shows why staging helps:
phases other than weight opening are already small under cache hits, so the
switch path is dominated by Phase~2. Full D2D staging reduces the p50
Phase~2 cost from 2.57\,s to 110\,ms, while partial staging lands between the
two according to the uncovered H2D fraction.

\paragraph{Path classification.}
Each rank-level record also identifies the load path, staged-byte coverage,
dirty staging bytes, and D2D/H2D gather times. In the TP=8 staged run used for
the first two path groups of Table~\ref{tab:switch-phases}, the filtered set contains
15{,}070 rank-level records: 14{,}952 D2D-full, 102 partial D2D+H2D, and 16
full-H2D fallbacks. These records hit the cached KV shell and CUDA graph, so the
tail is driven by uncovered weight bytes rather than runtime recapture. D2D-full
switches have a p50 D2D gather-sync time of 18.3\,ms; partial switches add H2D
synchronization with a p50 of 696.2\,ms for uncovered chunks. Dirty chunks arise
when active KV allocations overlap the staging range, and the guard reloads
those chunks from host rather than copying unsafe D2D sources.

\paragraph{Model-specific weight postprocessing.}
Some models expose runtime weights that are not a byte-for-byte copy of the raw
checkpoint tensors. In our GLM runs, the quantized post-load path materializes
packed and derived tensors on the GPU after the first open. The switch path
therefore treats the postprocessed physical layout as the stable representation:
after finalization, it writes the processed tensors back to the host-side model
entry and uses that layout for later H2D or D2D reloads. This keeps region
sizing and staged-byte accounting aligned with the physical parameter and
persistent-buffer order used by the active model.

\section{Policy Knobs}
\label{app:knobs}

\textbf{Reservation timeout $\tau$.}
We select $\tau=1$\,s from the sensitivity sweep in
\S\ref{sec:eval-reservation}; it is near the p90 inter-call gap in the
characterization trace.
Expired sessions fall back to normal residency-aware placement.

\textbf{Bump memory fraction.}
The component experiments use \texttt{mem-fraction-bump=0.8}, reserving most
HBM for the managed layout while leaving headroom for runtime allocations and a
safety margin.

\textbf{Admission margin $\gamma$.}
Cold-pool rounds are bounded below the configured TTFT target with
$\gamma<1$, after accounting for switching and restore overhead. This knob is
an admission-control margin; absolute target attainment still depends on the
offered load and provisioning level.

\textbf{Replica penalty $\nu_m$.}
The router uses a per-model replica penalty to avoid opening replicas for
transient bursts whose benefit does not cover switch and memory cost.

%% file: ref.bib
@inproceedings{shahrad2020serverless,
  author    = {Shahrad, Mohammad and Fonseca, Rodrigo and Goiri, Inigo and Chaudhry, Gohar and Batum, Paul and Cooke, Jason and Laureano, Eduardo and Tresness, Colby and Russinovich, Mark and Bianchini, Ricardo},
  title     = {Serverless in the Wild: Characterizing and Optimizing the Serverless Workload at a Large Cloud Provider},
  booktitle = {Proceedings of the 2020 USENIX Annual Technical Conference (USENIX ATC)},
  year      = {2020},
}

@inproceedings{crankshaw2017clipper,
  author    = {Crankshaw, Daniel and Wang, Xin and Zhou, Guilio and Franklin, Michael J. and Gonzalez, Joseph E. and Stoica, Ion},
  title     = {Clipper: A Low-Latency Online Prediction Serving System},
  booktitle = {Proceedings of the 14th USENIX Symposium on Networked Systems Design and Implementation (NSDI)},
  year      = {2017},
}

@inproceedings{gujarati2020serving,
  author    = {Gujarati, Arpan and Karimi, Reza and Alzayat, Safya and Hao, Wei and Kaufmann, Antoine and Vigfusson, Ymir and Mace, Jonathan},
  title     = {Serving {DNN}s Like Clockwork: Performance Predictability from the Bottom Up},
  booktitle = {Proceedings of the 14th USENIX Symposium on Operating Systems Design and Implementation (OSDI)},
  year      = {2020},
}

@misc{olston2017tensorflowserving,
  author        = {Olston, Christopher and Fiedel, Noah and Gorovoy, Kiril and Harmsen, Jeremiah and Lao, Li and Li, Fangwei and Rajashekhar, Vinu and Ramesh, Sukriti and Soyke, Jordan},
  title         = {{TensorFlow-Serving}: Flexible, High-Performance {ML} Serving},
  year          = {2017},
  eprint        = {1712.06139},
  archivePrefix = {arXiv},
  primaryClass  = {cs.DC},
}

@inproceedings{shen2019nexus,
  author    = {Shen, Haichen and Chen, Lequn and Jin, Yuchen and Zhao, Liangyu and Kong, Bingyu and Philipose, Matthai and Krishnamurthy, Arvind and Sundaram, Ravi},
  title     = {{Nexus}: A {GPU} Cluster Engine for Accelerating {DNN}-Based Video Analysis},
  booktitle = {Proceedings of the 27th ACM Symposium on Operating Systems Principles (SOSP)},
  year      = {2019},
  pages     = {322--337},
}

@inproceedings{crankshaw2020inferline,
  author    = {Crankshaw, Daniel and Sela, Gur-Eyal and Zumar, Corey and Mo, Xiangxi and Gonzalez, Joseph E. and Stoica, Ion and Tumanov, Alexey},
  title     = {{InferLine}: Latency-Aware Provisioning and Scaling for Prediction Serving Pipelines},
  booktitle = {Proceedings of the 11th ACM Symposium on Cloud Computing (SoCC)},
  year      = {2020},
}

@inproceedings{romero2021infaas,
  author    = {Romero, Francisco and Li, Qian and Yadwadkar, Neeraja J. and Kozyrakis, Christos},
  title     = {{INFaaS}: Automated Model-less Inference Serving},
  booktitle = {Proceedings of the 2021 USENIX Annual Technical Conference (USENIX ATC)},
  year      = {2021},
  pages     = {397--411},
}

@inproceedings{yu2022orca,
  author    = {Yu, Gyeong-In and Jeong, Joo Seong and Kim, Geon-Woo and Kim, Soojeong and Chun, Byung-Gon},
  title     = {Orca: A Distributed Serving System for Transformer-Based Generative Models},
  booktitle = {Proceedings of the 16th USENIX Symposium on Operating Systems Design and Implementation (OSDI)},
  year      = {2022},
}

@inproceedings{kwon2023vllm,
  author    = {Kwon, Woosuk and Li, Zhuohan and Zhuang, Siyuan and Sheng, Ying and Zheng, Lianmin and Yu, Cody Hao and Gonzalez, Joseph E. and Zhang, Hao and Stoica, Ion},
  title     = {Efficient Memory Management for Large Language Model Serving with PagedAttention},
  booktitle = {Proceedings of the ACM SIGOPS 29th Symposium on Operating Systems Principles (SOSP)},
  year      = {2023},
}

@inproceedings{fu2024serverlessllm,
  author    = {Fu, Yao and Xue, Leyang and Huang, Yeqi and Brabete, Andrei-Octavian and Ustiugov, Dmitrii and Patel, Yuvraj and Mai, Luo},
  title     = {{ServerlessLLM}: Low-Latency Serverless Inference for Large Language Models},
  booktitle = {Proceedings of the 18th USENIX Symposium on Operating Systems Design and Implementation (OSDI)},
  year      = {2024},
}

@inproceedings{zhong2024distserve,
  author    = {Zhong, Yinmin and Liu, Shengyu and Chen, Junda and Hu, Jianbo and Zhu, Yibo and Liu, Xuanzhe and Jin, Xin and Zhang, Hao},
  title     = {{DistServe}: Disaggregating Prefill and Decoding for Goodput-optimized Large Language Model Serving},
  booktitle = {Proceedings of the 18th USENIX Symposium on Operating Systems Design and Implementation (OSDI)},
  year      = {2024},
}

@inproceedings{agrawal2024sarathi,
  author    = {Agrawal, Amey and Kedia, Nitin and Panwar, Ashish and Mohan, Jayashree and Kwatra, Nipun and Gulavani, Bhargav S. and Tumanov, Alexey and Ramjee, Ramachandran},
  title     = {Taming {Throughput-Latency} Tradeoff in {LLM} Inference with {Sarathi-Serve}},
  booktitle = {Proceedings of the 18th USENIX Symposium on Operating Systems Design and Implementation (OSDI)},
  year      = {2024},
}

@inproceedings{patel2024splitwise,
  author    = {Patel, Pratyush and Choukse, Esha and Zhang, Chaojie and Shah, Aashaka and Goiri, I{\~n}igo and Maleki, Saeed and Bianchini, Ricardo},
  title     = {{Splitwise}: Efficient Generative {LLM} Inference Using Phase Splitting},
  booktitle = {Proceedings of the 51st Annual International Symposium on Computer Architecture (ISCA)},
  year      = {2024},
}

@inproceedings{wu2023fastserve,
  author    = {Wu, Bingyang and Zhong, Yinmin and Zhang, Zili and Huang, Gang and Liu, Xuanzhe and Jin, Xin},
  title     = {{FastServe}: Fast Distributed Inference Serving for Large Language Models},
  booktitle = {Proceedings of the 17th USENIX Symposium on Operating Systems Design and Implementation (OSDI)},
  year      = {2023},
}

@misc{holmes2024deepspeedfastgen,
  author        = {Holmes, Connor and Tanaka, Masahiro and Wyatt, Michael and Awan, Ammar Ahmad and Rasley, Jeff and Rajbhandari, Samyam and Aminabadi, Reza Yazdani and Qin, Heyang and Bakhtiari, Arash and Kurilenko, Lev and He, Yuxiong},
  title         = {{DeepSpeed-FastGen}: High-throughput Text Generation for {LLM}s via Dynamic SplitFuse},
  year          = {2024},
  eprint        = {2401.08671},
  archivePrefix = {arXiv},
  primaryClass  = {cs.LG},
}

@misc{lai2025tokenscale,
  author        = {Lai, Ruiqi and Liu, Hongrui and Lu, Chengzhi and Liu, Zonghao and Cao, Siyu and Shao, Siyang and Zhang, Yixin and Mai, Luo and Ustiugov, Dmitrii},
  title         = {{TokenScale}: Timely and Accurate Autoscaling for Disaggregated {LLM} Serving with Token Velocity},
  year          = {2025},
  eprint        = {2512.03416},
  archivePrefix = {arXiv},
  primaryClass  = {cs.DC},
}

@inproceedings{li2023alpaserve,
  author    = {Li, Zhuohan and Zheng, Lianmin and Zhong, Yinmin and Liu, Vincent and Sheng, Ying and Jin, Xin and Huang, Yanping and Chen, Zhifeng and Zhang, Hao and Gonzalez, Joseph E. and Stoica, Ion},
  title     = {{AlpaServe}: Statistical Multiplexing with Model Parallelism for Deep Learning Serving},
  booktitle = {Proceedings of the 17th USENIX Symposium on Operating Systems Design and Implementation (OSDI)},
  year      = {2023},
}

@inproceedings{xiang2025aegaeon,
  author    = {Xiang, Yuxing and Li, Xue and Qian, Kun and Yang, Yufan and Zhu, Diwen and Yu, Wenyuan and Zhai, Ennan and Liu, Xuanzhe and Jin, Xin and Zhou, Jingren},
  title     = {Aegaeon: Effective GPU Pooling for Concurrent LLM Serving on the Market},
  booktitle = {Proceedings of the ACM SIGOPS 31st Symposium on Operating Systems Principles (SOSP)},
  year      = {2025},
  doi       = {10.1145/3731569.3764815},
}

@misc{duan2024muxserve,
  author        = {Duan, Jiangfei and Lu, Runyu and Duanmu, Haojie and Li, Xiuhong and Zhang, Xingcheng and Lin, Dahua and Stoica, Ion and Zhang, Hao},
  title         = {{MuxServe}: Flexible Spatial-Temporal Multiplexing for Multiple {LLM} Serving},
  year          = {2024},
  eprint        = {2404.02015},
  archivePrefix = {arXiv},
  primaryClass  = {cs.DC},
}

@inproceedings{sheng2024slora,
  author    = {Sheng, Ying and Cao, Shiyi and Li, Dacheng and Hooper, Coleman and Lee, Nicholas and Yang, Shuo and Chou, Christopher and Zhu, Banghua and Zheng, Lianmin and Keutzer, Kurt and Gonzalez, Joseph E. and Stoica, Ion},
  title     = {{S-LoRA}: Serving Thousands of Concurrent {LoRA} Adapters},
  booktitle = {Proceedings of Machine Learning and Systems (MLSys)},
  year      = {2024},
}

@inproceedings{chen2024punica,
  author    = {Chen, Lequn and Ye, Zhiqiang and Wu, Yilong and Zhuo, Danyang and Ceze, Luis and Krishnamurthy, Arvind and Zhang, Tianqi},
  title     = {{Punica}: Multi-Tenant {LoRA} Serving},
  booktitle = {Proceedings of Machine Learning and Systems (MLSys)},
  year      = {2024},
}

@inproceedings{wu2024dlora,
  author    = {Wu, Bingyang and Zhu, Ruidong and Zhang, Zili and Sun, Peng and Liu, Xuanzhe and Jin, Xin},
  title     = {{dLoRA}: Dynamically Orchestrating Requests and Adapters for {LoRA} {LLM} Serving},
  booktitle = {Proceedings of the 18th USENIX Symposium on Operating Systems Design and Implementation (OSDI)},
  year      = {2024},
  pages     = {911--927},
}

@inproceedings{miao2024spotserve,
  author    = {Miao, Xupeng and Shi, Chunan and Duan, Jiangfei and Xi, Xiaoli and Lin, Dahua and Cui, Bin and Jia, Zhihao},
  title     = {{SpotServe}: Serving Generative Large Language Models on Preemptible Instances},
  booktitle = {Proceedings of the ACM International Conference on Architectural Support for Programming Languages and Operating Systems (ASPLOS)},
  year      = {2024},
}

@inproceedings{gao2024cachedattention,
  author    = {Gao, Bin and He, Zhuomin and Sharma, Puru and Kang, Qingxuan and Jevdjic, Djordje and Deng, Junbo and Yang, Xingkun and Yu, Zhou and Zuo, Pengfei},
  title     = {Cost-Efficient Large Language Model Serving for Multi-turn Conversations with {CachedAttention}},
  booktitle = {Proceedings of the 2024 USENIX Annual Technical Conference (USENIX ATC)},
  year      = {2024},
}

@inproceedings{sheng2023flexgen,
  author    = {Sheng, Ying and Zheng, Lianmin and Yuan, Binhang and Li, Zhuohan and Ryabinin, Max and Fu, Daniel Y. and Xie, Zhiqiang and Chen, Beidi and Barrett, Clark and Gonzalez, Joseph E. and Liang, Percy and R{\'e}, Christopher and Stoica, Ion and Zhang, Ce},
  title     = {{FlexGen}: High-Throughput Generative Inference of Large Language Models with a Single {GPU}},
  booktitle = {Proceedings of the 40th International Conference on Machine Learning (ICML)},
  year      = {2023},
}

@inproceedings{zheng2024sglang,
  author    = {Zheng, Lianmin and Yin, Liangsheng and Xie, Zhiqiang and Sun, Chuyue and Huang, Jeff and Yu, Cody Hao and Cao, Shiyi and Kozyrakis, Christos and Stoica, Ion and Gonzalez, Joseph E. and Barrett, Clark and Sheng, Ying},
  title     = {{SGLang}: Efficient Execution of Structured Language Model Programs},
  booktitle = {Advances in Neural Information Processing Systems (NeurIPS)},
  year      = {2024},
}

@misc{wu2024loongserve,
  author        = {Wu, Bingyang and Liu, Shengyu and Zhong, Yinmin and Sun, Peng and Liu, Xuanzhe and Jin, Xin},
  title         = {{LoongServe}: Efficiently Serving Long-Context Large Language Models with Elastic Sequence Parallelism},
  year          = {2024},
  eprint        = {2404.09526},
  archivePrefix = {arXiv},
  primaryClass  = {cs.DC},
}

@inproceedings{lee2024infinigen,
  author    = {Lee, Wonbeom and Lee, Jungi and Seo, Junghwan and Sim, Jaewoong},
  title     = {{InfiniGen}: Efficient Generative Inference of Large Language Models with Dynamic {KV} Cache Management},
  booktitle = {Proceedings of the 18th USENIX Symposium on Operating Systems Design and Implementation (OSDI)},
  year      = {2024},
}

@inproceedings{li2024cachegen,
  author    = {Liu, Yuhan and Li, Hanchen and Cheng, Yihua and Ray, Siddhant and Huang, Yuyang and Zhang, Qizheng and Du, Kuntai and Yao, Jiayi and Lu, Shan and Ananthanarayanan, Ganesh and Maire, Michael and Hoffmann, Henry and Holtzman, Ari and Jiang, Junchen},
  title     = {{CacheGen}: {KV} Cache Compression and Streaming for Fast Large Language Model Serving},
  booktitle = {Proceedings of the ACM SIGCOMM Conference},
  year      = {2024},
}

@misc{hu2024memserve,
  author        = {Hu, Cunchen and Huang, Heyang and Hu, Junhao and Xu, Jiang and Chen, Xusheng and Xie, Tao and Wang, Chenxi and Wang, Sa and Bao, Yungang and Sun, Ninghui and Shan, Yizhou},
  title         = {{MemServe}: Context Caching for Disaggregated {LLM} Serving with Elastic Memory Pool},
  year          = {2024},
  eprint        = {2406.17565},
  archivePrefix = {arXiv},
  primaryClass  = {cs.DC},
}

@misc{hu2024tetriinfer,
  author        = {Hu, Cunchen and Huang, Heyang and Xu, Liangliang and Chen, Xusheng and Xu, Jiang and Chen, Shuang and Feng, Hao and Wang, Chenxi and Wang, Sa and Bao, Yungang and Sun, Ninghui and Shan, Yizhou},
  title         = {Inference without Interference: Disaggregate {LLM} Inference for Mixed Downstream Workloads},
  year          = {2024},
  eprint        = {2401.11181},
  archivePrefix = {arXiv},
  primaryClass  = {cs.DC},
}

@misc{liu2024lmcache,
  author        = {Cheng, Yihua and Liu, Yuhan and Yao, Jiayi and An, Yuwei and Chen, Xiaokun and Feng, Shaoting and Huang, Yuyang and Shen, Samuel and Du, Kuntai and Jiang, Junchen},
  title         = {{LMCache}: An Efficient {KV} Cache Layer for Enterprise-Scale {LLM} Inference},
  year          = {2025},
  eprint        = {2510.09665},
  archivePrefix = {arXiv},
  primaryClass  = {cs.DC},
}

@inproceedings{qin2025mooncake,
  author    = {Qin, Ruoyu and Li, Zheming and He, Weiran and Cui, Jialei and Tang, Heyi and Ren, Feng and Ma, Teng and Cai, Shangming and Zhang, Yineng and Zhang, Mingxing and Wu, Yongwei and Zheng, Weimin and Xu, Xinran},
  title     = {Mooncake: Trading More Storage for Less Computation -- A {KVCache}-centric Architecture for Serving {LLM} Chatbot},
  booktitle = {Proceedings of the 23rd USENIX Conference on File and Storage Technologies (FAST)},
  year      = {2025},
}

@misc{srivatsa2024preble,
  author        = {Srivatsa, Vikranth and He, Zijian and Abhyankar, Reyna and Li, Dongming and Zhang, Yiying},
  title         = {{Preble}: Efficient Distributed Prompt Scheduling for {LLM} Serving},
  year          = {2024},
  eprint        = {2407.00023},
  archivePrefix = {arXiv},
  primaryClass  = {cs.DC},
}

@misc{yuan2026dualmap,
  author        = {Yuan, Ying and Zuo, Pengfei and Wang, Bo and Chen, Zhangyu and Tan, Zhipeng and Yu, Zhou},
  title         = {{DualMap}: Enabling Both Cache Affinity and Load Balancing for Distributed {LLM} Serving},
  year          = {2026},
  eprint        = {2602.06502},
  archivePrefix = {arXiv},
  primaryClass  = {cs.DC},
}

@inproceedings{wu2024llumnix,
  author    = {Sun, Biao and Huang, Ziming and Zhao, Hanyu and Xiao, Wencong and Zhang, Xinyi and Li, Yong and Lin, Wei},
  title     = {{Llumnix}: Dynamic Scheduling for Large Language Model Serving},
  booktitle = {Proceedings of the 18th USENIX Symposium on Operating Systems Design and Implementation (OSDI)},
  year      = {2024},
}

@misc{luo2025autellix,
  author       = {Luo, Michael and Shi, Xiaoxiang and Cai, Colin and Zhang, Tianjun and Wong, Justin and Wang, Yichuan and Wang, Chi and Huang, Yanping and Chen, Zhifeng and Gonzalez, Joseph E. and Stoica, Ion},
  title        = {Autellix: An Efficient Serving Engine for LLM Agents as General Programs},
  year         = {2025},
  eprint       = {2502.13965},
  archivePrefix = {arXiv},
  primaryClass = {cs.LG},
}

@misc{li2025continuum,
  author       = {Li, Hanchen and He, Runyuan and Mang, Qiuyang and Zhang, Qizheng and Mao, Huanzhi and Chen, Xiaokun and Zhou, Hangrui and Cheung, Alvin and Gonzalez, Joseph E. and Stoica, Ion},
  title        = {Continuum: Efficient and Robust Multi-Turn LLM Agent Scheduling with KV Cache Time-to-Live},
  year         = {2025},
  eprint       = {2511.02230},
  archivePrefix = {arXiv},
  primaryClass = {cs.OS},
}

@inproceedings{lin2024parrot,
  author    = {Lin, Chaofan and Han, Zhenhua and Zhang, Chengruidong and Yang, Yuqing and Yang, Fan and Chen, Chen and Qiu, Lili},
  title     = {Parrot: Efficient Serving of {LLM-based} Applications with Semantic Variable},
  booktitle = {Proceedings of the 18th USENIX Symposium on Operating Systems Design and Implementation (OSDI)},
  year      = {2024},
  pages     = {929--945},
}

@inproceedings{gim2025pie,
  author    = {Gim, In and Ma, Zhiyao and Lee, Seung-seob and Zhong, Lin},
  title     = {{Pie}: A Programmable Serving System for Emerging {LLM} Applications},
  booktitle = {Proceedings of the ACM SIGOPS 31st Symposium on Operating Systems Principles (SOSP)},
  year      = {2025},
}

@misc{ghosh2026hfopensource,
  author       = {Ghosh, Avijit and Kaffee, Lucie-Aim{\'e}e and Jernite, Yacine and Solaiman, Irene},
  title        = {State of Open Source on Hugging Face: Spring 2026},
  howpublished = {Hugging Face Blog},
  year         = {2026},
  url          = {https://huggingface.co/blog/huggingface/state-of-os-hf-spring-2026},
}

@techreport{stanford2025aiindex,
  author      = {{Stanford Institute for Human-Centered Artificial Intelligence}},
  title       = {The 2025 AI Index Report},
  institution = {Stanford University},
  year        = {2025},
  url         = {https://hai.stanford.edu/ai-index/2025-ai-index-report},
}

@inproceedings{jimenez2024swebench,
  author    = {Jimenez, Carlos E. and Yang, John and Wettig, Alexander and Yao, Shunyu and Pei, Kexin and Press, Ofir and Narasimhan, Karthik R.},
  title     = {{SWE}-bench: Can Language Models Resolve Real-World {GitHub} Issues?},
  booktitle = {Proceedings of the 12th International Conference on Learning Representations (ICLR)},
  year      = {2024},
}

@inproceedings{yao2023react,
  author    = {Yao, Shunyu and Zhao, Jeffrey and Yu, Dian and Du, Nan and Shafran, Izhak and Narasimhan, Karthik and Cao, Yuan},
  title     = {{ReAct}: Synergizing Reasoning and Acting in Language Models},
  booktitle = {Proceedings of the 11th International Conference on Learning Representations (ICLR)},
  year      = {2023},
}

@inproceedings{schick2023toolformer,
  author    = {Schick, Timo and Dwivedi-Yu, Jane and Dess{\`i}, Roberto and Raileanu, Roberta and Lomeli, Maria and Zettlemoyer, Luke and Cancedda, Nicola and Scialom, Thomas},
  title     = {{Toolformer}: Language Models Can Teach Themselves to Use Tools},
  booktitle = {Advances in Neural Information Processing Systems (NeurIPS)},
  year      = {2023},
}

@inproceedings{yang2024sweagent,
  author    = {Yang, John and Jimenez, Carlos E. and Wettig, Alexander and Lieret, Kilian and Yao, Shunyu and Narasimhan, Karthik and Press, Ofir},
  title     = {{SWE}-agent: Agent-Computer Interfaces Enable Automated Software Engineering},
  booktitle = {Advances in Neural Information Processing Systems (NeurIPS)},
  year      = {2024},
}

@misc{prabhu2024vattention,
  author        = {Prabhu, Ramya and Nayak, Ajay and Mohan, Jayashree and Ramjee, Ramachandran and Panwar, Ashish},
  title         = {{vAttention}: Dynamic Memory Management for Serving {LLM}s without {PagedAttention}},
  year          = {2024},
  eprint        = {2405.04437},
  archivePrefix = {arXiv},
  primaryClass  = {cs.LG},
}

@misc{lin2024infinitellm,
  author        = {Lin, Bin and Zhang, Chen and Peng, Tao and Zhao, Hanyu and Xiao, Wencong and Sun, Minmin and Liu, Anmin and Zhang, Zhipeng and Li, Lanbo and Qiu, Xiafei and Li, Shen and Ji, Zhigang and Xie, Tao and Li, Yong and Lin, Wei},
  title         = {Infinite-{LLM}: Efficient {LLM} Service for Long Context with {DistAttention} and Distributed {KVCache}},
  year          = {2024},
  eprint        = {2401.02669},
  archivePrefix = {arXiv},
  primaryClass  = {cs.DC},
}

@misc{juravsky2024hydragen,
  author        = {Juravsky, Jordan and Brown, Bradley and Ehrlich, Ryan and Fu, Daniel Y. and R{\'e}, Christopher and Mirhoseini, Azalia},
  title         = {{Hydragen}: High-Throughput {LLM} Inference with Shared Prefixes},
  year          = {2024},
  eprint        = {2402.05099},
  archivePrefix = {arXiv},
  primaryClass  = {cs.LG},
}

@misc{yao2024cacheblend,
  author        = {Yao, Jiayi and Li, Hanchen and Liu, Yuhan and Ray, Siddhant and Cheng, Yihua and Zhang, Qizheng and Du, Kuntai and Lu, Shan and Jiang, Junchen},
  title         = {{CacheBlend}: Fast Large Language Model Serving for {RAG} with Cached Knowledge Fusion},
  year          = {2024},
  eprint        = {2405.16444},
  archivePrefix = {arXiv},
  primaryClass  = {cs.LG},
}

@misc{zhu2024nanoflow,
  author        = {Zhu, Kan and Zhao, Yilong and Zhao, Liang and Zuo, Gan and Gu, Yile and Xie, Dedong and Gao, Yufei and Xu, Qinyu and Tang, Tian and Ye, Zihao and Kamahori, Keisuke and Lin, Chien-Yu and Wang, Stephanie and Krishnamurthy, Arvind and Kasikci, Baris},
  title         = {{NanoFlow}: Towards Optimal Large Language Model Serving Throughput},
  year          = {2024},
  eprint        = {2408.12757},
  archivePrefix = {arXiv},
  primaryClass  = {cs.DC},
}

@misc{xu2023rewoo,
  author        = {Xu, Binfeng and Peng, Zhendong and Lei, Bowen and Mukherjee, Subhabrata and Liu, Yu and Xu, Dongkuan},
  title         = {{ReWOO}: Decoupling Reasoning from Observations for Efficient Augmented Language Models},
  year          = {2023},
  eprint        = {2305.18323},
  archivePrefix = {arXiv},
  primaryClass  = {cs.CL},
}

@inproceedings{shinn2023reflexion,
  author    = {Shinn, Noah and Cassano, Federico and Gopinath, Ashwin and Narasimhan, Karthik and Yao, Shunyu},
  title     = {Reflexion: Language Agents with Verbal Reinforcement Learning},
  booktitle = {Advances in Neural Information Processing Systems (NeurIPS)},
  year      = {2023},
}

@misc{wu2023autogen,
  author        = {Wu, Qingyun and Bansal, Gagan and Zhang, Jieyu and Wu, Yiran and Li, Beibin and Zhu, Erkang and Jiang, Li and Zhang, Xiaoyun and Zhang, Shaokun and Liu, Jiale and Awadallah, Ahmed Hassan and White, Ryen W. and Burger, Doug and Wang, Chi},
  title         = {{AutoGen}: Enabling Next-Gen {LLM} Applications via Multi-Agent Conversation},
  year          = {2023},
  eprint        = {2308.08155},
  archivePrefix = {arXiv},
  primaryClass  = {cs.AI},
}
